\documentclass[12pt]{iopart}
\usepackage{epsfig,graphicx,multicol,bbm,mathrsfs,amssymb,dcolumn,bm}
 \usepackage{cite}

\begin{document}

\title[Low energy e/recoil discrimination for directional Dark Matter detection]{Low energy electron/recoil discrimination for directional Dark Matter detection}
\author{J. Billard, F. Mayet and D. Santos}
\address{Laboratoire de Physique Subatomique et de Cosmologie, Universit\'e Joseph Fourier Grenoble 1,
  CNRS/IN2P3, Institut Polytechnique de Grenoble, 53 rue des Martyrs, Grenoble, France }
\ead{billard@lpsc.in2p3.fr}

\begin{abstract}
Directional detection is a promising Dark Matter search strategy. 
Even though it could accommodate to a sizeable background contamination, electron/recoil discrimination 
remains a key and challenging issue as for direction-insensitive detectors. The measurement of the 3D track may be 
used to discriminate electrons from nuclear recoils. While a high rejection power is expected above 20 keV ionization, a dedicated data
analysis is needed at low energy. After identifying discriminant observables, a multivariate analysis, namely a Boosted Decision Tree, is 
proposed, enabling an efficient event tagging for Dark Matter search. We show that it allows us to optimize  rejection while keeping 
a rather high efficiency which is compulsory for rare event search.With respect to a  sequential analysis, the 
rejection is about $\sim 20$ times higher with a multivariate analysis, for the same Dark Matter exclusion limit.
 \end{abstract}

\pacs{95.35.+d, 29.40.Cs, 29.85.Fj}
\maketitle


Directional detection of Dark Matter has been first proposed as a powerful tool to
identify genuine WIMP events as such~\cite{spergel}. Since then, the contribution of 
directional detection to the field of Dark Matter has been adressed through a wealth of 
studies~\cite{albornoz,Alves:2012ay,billard.disco,billard.ident,billard.profile,billard.exclusion,Bozorgnia:2011vc,Chiang:2012ze,copi1,copi2,copi3,Creswick:2010dm,Kuhlen:2012fz,Lee:2012pf,Lisanti:2009vy,
green1,green2,green.disco,morgan1,morgan2}. In particular, it has been shown that even a low exposure directional detector (30 kg.year) could lead to  
a high significance discovery of galactic Dark Matter \cite{billard.disco,billard.profile,green.disco}. For neutralino Dark Matter, most MSSM configurations with a neutralino lighter
than 200 $\rm GeV/c^2$ would lead to a significance greater than 3$\sigma$
(90\% CL) in a 30 kg.year CF4 directional detector \cite{albornoz}.\\
There is a worldwide effort toward the development of a large directional detector 
\cite{white,dmtpc,drift,d3,mimac,newage}, most projects being low pressure Time Projection Chambers (TPC).  All current projects face common challenges, 
some being specific to directional detection, some being shared with direction-insensitive strategy \cite{censier}. In particular, zero background is often referred to as the ultimate goal 
for the next generation of direct detection experiments in deep underground laboratories. Even though directional detection could accommodate to a sizeable background 
contamination  \cite{billard.profile,billard.exclusion},  owing to
the large intrinsic difference between the WIMP-induced and background-induced
spectra, electron/recoil discrimination remains a key and challenging issue as for 
direction-insensitive detectors. For the latter, electron/recoil discrimination may be achieved 
thanks to a  double detection of the energy :  heat/ionization \cite{cdms,edelweiss}, scintillation/ionization \cite{xenon1,xenon2} 
or heat/scintillation \cite{cresst}. As far as directional detection is concerned, only the ionization energy is measured and the
discrimination cannot rely on the measurement of the energy deposit only. However, as
early suggested in \cite{moulin.these,Santos:2005xj}, the information on the 3D track may be used to discriminate electron and nuclear
recoils. A recent study \cite{billard.track} has shown the possibility to achieve 3D reconstruction in the case of the MIMAC 
project~\cite{mimac}. In particular, the vertex (X, Y and Z) of the elastic scatterring interaction can be retrieved, thus enabling a 
fiducialization of the detector and hence reducing the background contamination from surface events. However an  e/recoil
discrimination method would be a powerful way to further reduce background contamination.\\
In this paper, we propose a new method to discriminate electron from nuclear recoils. It is based on discriminant observables related to
track topology. High rejection power is expected at 
high energy (above 50 keV recoil) but low energy discrimination requires a careful data analysis and a 
multivariate analysis, namely a Boosted Decision Tree, is  proposed to optimize  rejection while keeping a 
rather high efficiency.  This event tagging based on a  Boosted Decision Tree method  is of general interest for Dark Matter search and could be
adapted for various e/recoil discrimination strategies.\\

The paper is organised as follows, we first choose discriminant observables and study their correlations 
(sec.~\ref{sec:obs}). Then, an electron/recoil discrimination is proposed within the framework of a sequential analysis 
(sec.~\ref{sec:oldschool}). In section \ref{sec:newschool}, we introduce a Boosted Decision Tree method, used to tag events either as
electron or nuclear recoil. We conclude in sec.~\ref{sec:con} by showing the consequences in terms of Dark Matter search.

\section{Discriminant observables}
\label{sec:obs}
In the following, we present the MIMAC track reconstruction strategy 
(sec.~\ref{sec:mimac}), discuss a full simulation of tracks in the MIMAC detector (sec.~\ref{sec:simu}). Then,  
we  choose discriminant observables (sec.~\ref{sec:obs2}) and study their correlations (sec.~\ref{sec:obs3}).
\subsection{MIMAC track measurement strategy}
\label{sec:mimac}
The MIMAC detector is a project dedicated to directional Dark Matter search \cite{mimac}. It is a matrix design of elementary $CF_4$ TPC operated at low pressure (50 mbar).
The primary electron-ion pairs produced by a nuclear recoil in one chamber of the matrix are detected by drifting 
the primary electrons to the grid of a bulk Micromegas \cite{Giomataris1,Iguaz:2011yc} and producing the avalanche 
in a very thin gap (128 or 256 $\mu$m).  The electrons move towards the grid in the drift space and are projected on the pixelized anode thus allowing to get 
information on the X and Y coordinates. A bulk Micromegas \cite{Giomataris1,Iguaz:2011yc}  with a 10.8 by 10.8 cm$^2$ active area, 
segmented in pixels with an effective pitch of 424 $\mu$m is used as 2D readout. In order to reconstruct the third dimension of the recoil, the Z coordinate {\it i.e.} along the drift axis, 
 a self-triggered electronics has been developed \cite{Richer:2009pi,Bourrion1,Bourrion2}.
The ionization energy measurement is done by using a charge integrator connected to the grid   sampled at a frequency of 
50 MHz. Then, to recover the kinetic energy of the recoiling nucleus, one has to know accurately the value of the Ionization Quenching 
Factor (IQF) \cite{guillaudin}. With such a measurement, the X and Y coordinates are measured on the anode, while the Z coordinate is retrieved from the 50 MHz
sampling of the anode. Hence, the track is 3D reconstructed, providing the electron drift velocity is known \cite{billard.these,magboltz}. A dedicated data analysis
method \cite{billard.track} allows us to retrieve, for each track, the initial 
recoil direction ($\theta, \phi$) together with the vertex (X, Y and Z) of the WIMP elastic scatterring interaction. In particular, 
the latter  enables a detector fiducialization which reduces the background contamination from surface events. However an  e/recoil
discrimination method would be a powerful way to further reduce background contamination.

\subsection{Track simulation in MIMAC}
\label{sec:simu}
We use the same simulation software as in \cite{billard.track}. For the reader's convenience,  we recall  the main points hereafter, noticing that a comprehensive discussion on 
systematics of track measurement may be found in \cite{billard.track}. As a matter of fact, a realistic track simulation in a low pressure TPC should include all processes and systematics:  
generation of electron or nuclear recoils, primary electron generation, electron drift properties, readout and trigger strategy. The simulation scheme is the following : 
\begin{itemize}
\item Electron and nuclear recoils are simulated respectively with the Geant4 software \cite{g4}  and the SRIM software \cite{srim}. 
The kinetic energy spectrum is chosen as flat, between 0 and 50 keV for electrons and 1 and
120 keV for Fluorine recoils. All events are randomly generated within the whole detector volume ($10.8 \times 10.8 \times 25 \ {\rm
cm}^3$)  and with an isotropic distribution.
\item Electron/ion pair generation is accounted for.  
For a given energy loss, primary electrons are 
randomly generated between two collision points. 
\item Electron drift properties are estimated with the Magboltz software \cite{magboltz}, 
The drift velocity ($v_d$) and   both the transversal and longitudinal diffusion ($D_l$ and $D_t$) are accounted for. The gas mixture is 70\% CF$_4$ and 30\% de CHF$_3$ at 50 mbar. 
 For a 100 V/cm drift field, we found  : $v_d = 21.4$ $\mu$m/ns,  $D_l = 278.4 \ \mu {\rm m/ \sqrt{cm}}$ and $D_t = 246.0 \ \mu {\rm m/ \sqrt{cm}}$.
\item The Micromegas is simulated with a 100\% transparency and a gain of $\sim$ 10$^3$, ensuring that 
each row of pixels is sensitive to the unique electron. 
\item Eventually, the track enveloppe formed by primary electrons generated along the track is then sampled with the MIMAC strategy. The pixelized anode is composed of 256 channels in X and Y, with an
effective width of 424 $\mu$m. Each pixel is connected either with a X or Y channel. For a given event, an X-Y coincidence is thus required to locate the position on the anode. The Z coordinate is then reconstructed
thanks to the 50 MHz sampling of the anode. Hence, each track is characterized by  $N_c$ time coincidences.
\end{itemize} 
As the simulated tracks must contain all systematics associated to data tracks, we emphasize that  
the MIMAC readout and trigger strategy are accounted, which is a key point of the simulation scheme. This allows us to include 
time sampling, pitch size, thresholds, missing coincidences and experimental limitations..
This simulation software provides us with a realistic set of simulated tracks, with various energies, drift sizes or gas properties.\\
A loose selection is imposed on simulated events : all events must have at least one coincidence $N_c>1$ and be 
fully enclosed within the detector, {\it i.e.} no border strips fired.

\subsection{Defining discriminant observables}
\label{sec:obs2}
\begin{figure}[t]
\begin{center}
\includegraphics[scale=0.38,angle=0]{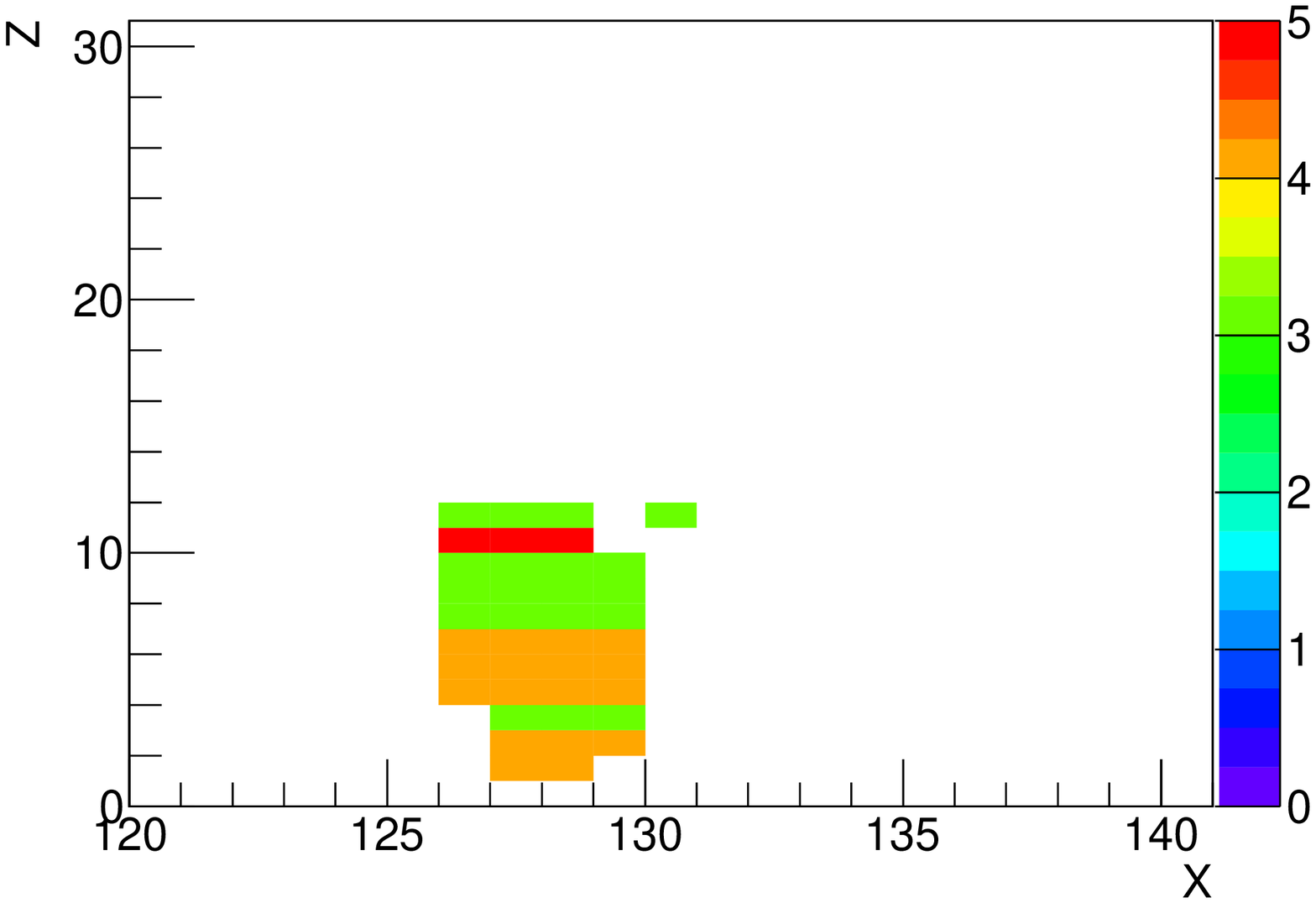}
\includegraphics[scale=0.38,angle=0]{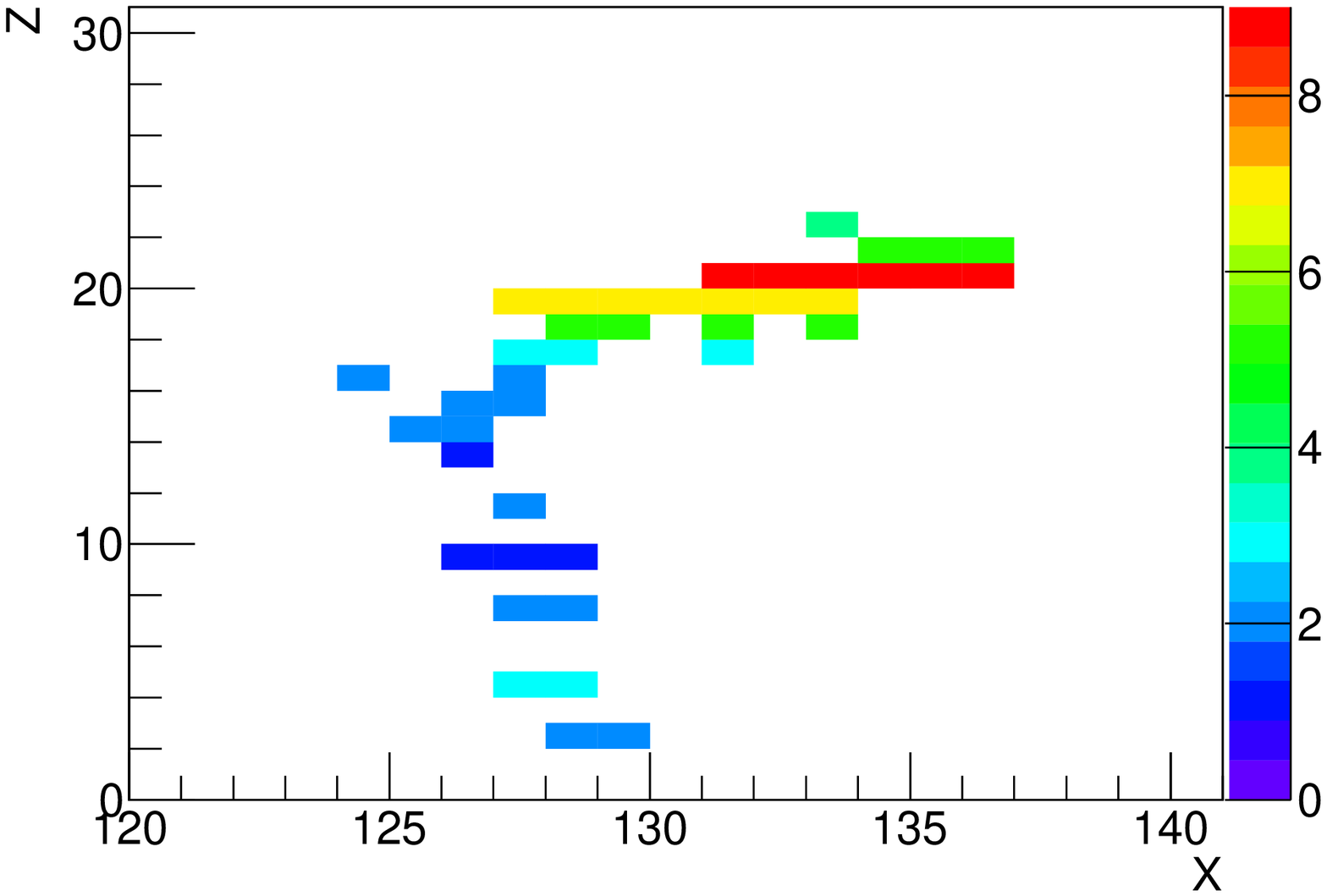}
\caption{2D projections, in the  (X,Z) plane, of simulated tracks :  a 100 keV Fluorine (left) and a 
10 keV electrons (right). See text for details.} 
\label{fig:PlanXZMIMAC}
\end{center}
\end{figure}

\begin{figure}[t]
\begin{center}
\includegraphics[scale=0.5,angle=0]{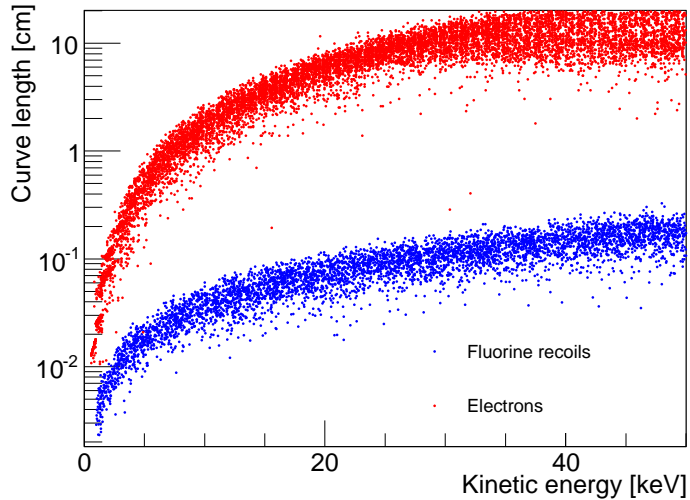}
\caption{Curve length as a function of the kinetic energy, for electrons (red) and Fluorine recoils (blue). 
Result obtained from raw tracks from the simulation,  before accounting for electron drift properties and detector sampling.} 
\label{fig:LVsLRaw}
\end{center}
\end{figure}

\begin{figure}[t]
\begin{center}
\includegraphics[scale=0.8,angle=0]{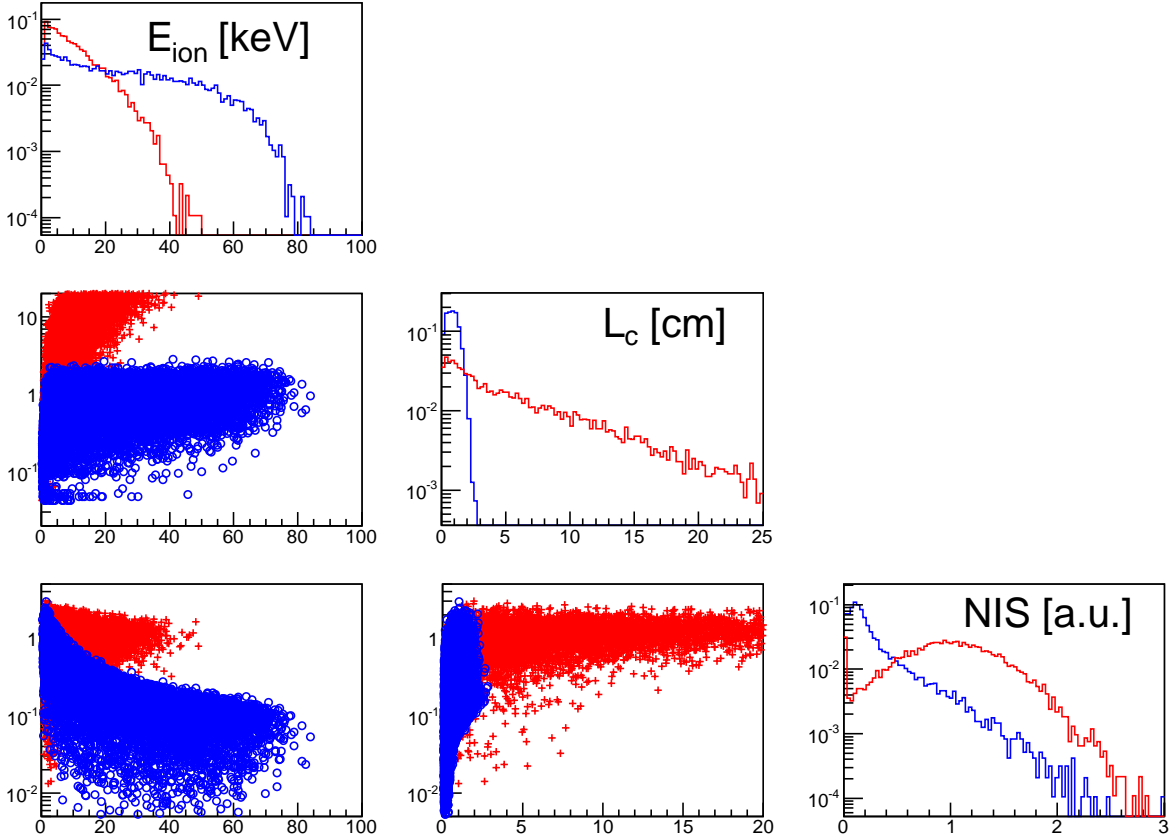}
\caption{Marginalised distributions (diagonal) and 2D correlations (off-diagonal) plots 
of the 3 observables ($E_{ion}$, $L_c$ and NIS). Electrons (red crosses) and Fluorine recoils (blue circles) are presented.}
\label{fig:FullDiscri}
\end{center}
\end{figure}

For a given energy, the track topology is expected to depend on the 
recoiling particle (electron or nucleus) and discriminant observables should be identified to evaluate the expected 
rejection  power and efficiency. Figure~\ref{fig:PlanXZMIMAC} presents the 2D projections, in the  (X,Z) plane, of simulated tracks :  a 100 keV Fluorine (left) and a 10 keV electrons (right). 
Both tracks are simulated with a common inital vertex (1 cm from the anode and in its center) and a common direction (perpendicular 
to the anode and going to the cathode). It is straightforward to notice that the topology of the tracks are very different. 
First,  the length of the electron track is much longer than the 
one of the nuclear recoil, even with a kinetic energy ten time smaller. Hence, the first idea is to use the track length as a 
discriminant observable, since for a given kinetic energy the range of electrons is expected to be much larger than 
the one of nuclear recoils. However, instead of using  
the projected range as an estimation of the track length, we define the curve length ($L_c$) as :   
\begin{equation}
L_c = \sum_{i = 1}^{N_c-1}\sqrt{(X_{i+1} - X_{i})^2 + (Y_{i+1} - Y_{i})^2 + (Z_{i+1} - Z_{i})^2}
\end{equation}
with $N_c$ the number of coincidences, ($X_i,Y_i,Z_i$) the barycentre location for each coincidence. $L_c$ is thus the sum of segments linking each consecutive barycentre.
The interest of this observable is the fact that it is strongly correlated to the range but it does also include the effect 
of small angular deflections along the track (see fig.~\ref{fig:PlanXZMIMAC}). This
enhances the difference between electrons and nuclear recoils and reduces the RMS of the range distribution to get a better 
discrimination. We highlight the fact that the curve length definition also includes electron longitudinal diffusion effect. 
For instance, it is about 1 mm for a 16 cm drift. Figure~\ref{fig:LVsLRaw} presents the curve length for simulated electrons (red) and
nuclear recoils (blue) as a function of the kinetic energy. This result has been obtained from raw tracks from the simulation, before
accounting for electron drift properties and detector sampling. It must hence be taken as the best discrimination  that can be obtained
from this observable (see fig.~\ref{fig:FullDiscri} to estimate the effect of drift and sampling). 
The ratio between the average electron curve length and the nuclear recoil 
one ranges between 1 and 2 orders of magnitude, depending on the kinetic energy. Note that above 30 keV,  
the electron distribution is getting broader due to the size of the detector ($10.8 \times 10.8 \times 25 \ {\rm cm}^3$).\\

The second observable is related to the track angular deviations. Indeed, as can be noticed 
on fig.~\ref{fig:PlanXZMIMAC}, a nuclear recoil track seem to be much more compact, with less angular deflections   
than an electron one. The observable includes the number of kinks encountered by the recoiling particle, and their magnitude. We define the Normalized Integrated Straggling (NIS) as : 
\begin{equation}
{\rm NIS} = \frac{1}{E_{ion}}\sum_{i=1}^{N_c-2}\Delta\theta_i
\end{equation}
It is the sum of angular deviation ($\Delta\theta_i$) between each consecutive barycentre 
($X_i, Y_i, Z_i$), normalized by the total ionization energy. The ionization energy is taken in arbitrary units in the definition as it
depends on the calibration in real life experiments.\\

Of course, other observables could be deduced from such 3D tracks, but they are all correlated with the one we have chosen and with a smaller discrimination power. For instance the projected range is closely
related to the curve length. The fraction of spatial coincidences  ($N_c/\Delta t_e$), close to 1 for nuclear recoils and ranging between 0 and 1 for electrons, is also correlated with the NIS and the curve
length. We converge to 3 observables ($E_{ion}, L_c$ and NIS) to discriminate electrons from nuclear recoils.

\subsection{Distribution and correlation of discriminant observables}
\label{sec:obs3}

 In the following, we study the 
distributions of these observables and their correlations in order to define 
the selections used in the following.
Figure~\ref{fig:FullDiscri}  presents the marginalised distributions (diagonal) and 2D correlations (off-diagonal) plots 
of the 3 observables. Electrons (red) and nuclear recoils (blue) are presented.\\ 

\noindent {\bf Ionization energy}\\
First, concerning the ionization energy (first row, first column), it is worth emphasizing  that the simulated 
recoil energy spectrum is chosen as flat, between 0 and 50 keV for electrons and between 1 and
120 keV for Fluorine recoils. The distribution of ionization energy of Fluorine recoils is steadily decreasing 
with an end point close to
80 keV. This is due to the ionization quenching effect. Indeed, at 120 keV the ionization quenching factor, from SRIM simulation, 
is    of the order of 65\%.  Concerning electron recoils, the ionization energy distribution
is also decreasing due to a finite-size effect. Indeed, at high energy the typical electron track length, whatever the definition, 
reaches the size of the detector. As the events are generated randomly within the detector volume, the effect appears also at lower energy. 
Note that there is of course no finite-size effect for nuclear recoils as the track length are very small, compared with the size of the
detector.\\

\noindent {\bf Curve length}\\
Figure~\ref{fig:FullDiscri} (second row, second column) presents the  marginalised distribution of the curve length.   
As expected, the curve length of nuclear recoils is very small, with a mean value of 1.5 cm and a maximum value of 3 cm, while electrons
present a broad distribution reaching values up to 20 cm. The electron curve length distribution presents the same finite-size effect as
  the energy distribution. Figure~\ref{fig:FullDiscri} (second row, first column) presents the distributions in the 
($L_c$, $E_{\rm ion}$) plane for electrons (red crosses) and Fluorine recoils (blue circles). The main feature to highlight is the fact
that the curve length value of Fluorine recoils does not depend on the ionization energy. This is due to the fact that for nuclear 
recoils (Fluorine in our case), the length is dominated by the electron diffusion. In fact, the detector is only weakly sensitive to the
recoil track length itself in this case and this effect is further enhanced by our definition of the track length ($L_c$). Note that
fluctuations on the X and Y barycentres also induces an oversetimation of the track length. 
These effects explain the difference in the mean curve length values of  fig. \ref{fig:LVsLRaw} and fig.~\ref{fig:FullDiscri}. On the contrary, 
for electrons the value of the curve length is  increasing with energy. Above 20 keV (ionization), the electron and nuclear recoil distributions are
well separated and even a straightforward analysis should allow to tag events with a high efficiency and high rejection power. The large
difference with respect to fig.~\ref{fig:LVsLRaw} highlights the need to include  all systematics associated to data tracks 
(electron drift,  readout, trigger strategy). Indeed, the e/recoil discrimination power reached {\it in fine} depends both on the quality
of the track measurement and of the data analysis strategy.\\

\noindent {\bf NIS observable}\\
As shown figure~\ref{fig:FullDiscri} (third row, third column), the NIS distribution of nuclear recoils remains close to 0, 
whereas the electron one presents a mean value of 1 (A.U.). We note
that the null values of NIS are related to the minimal number of coincidences required to define an angular deviation ($N_c \geq 2$). 
Figure~\ref{fig:FullDiscri} (third row, first column)   presents the distributions in the 
(NIS, $E_{\rm ion}$) plane. At low energy, below $\sim$ 20 keV, the two distributions are superimposed. Above 20 keV (ionization), the values
of the NIS observable of electrons and fluorine recoils start  being well separated. Contrarily to the $L_c$ behaviour, the value of NIS
does not depend on the energy for electrons and is  strongly decreasing with energy for nuclear recoils. For instance, at 50 keV, the 
NIS is about 10 time smaller than at 5 keV, highlighting the fact that high energy  Fluorine recoils are rectilinear tracks.\\

\noindent
This study show that these observables may be used to discriminate electrons from nuclear recoils. Before using them within the framework
of a dedicated data analysis, we check that they are not correlated with each other.\\

\begin{figure}[t]
\begin{center}
\includegraphics[scale=0.38,angle=0]{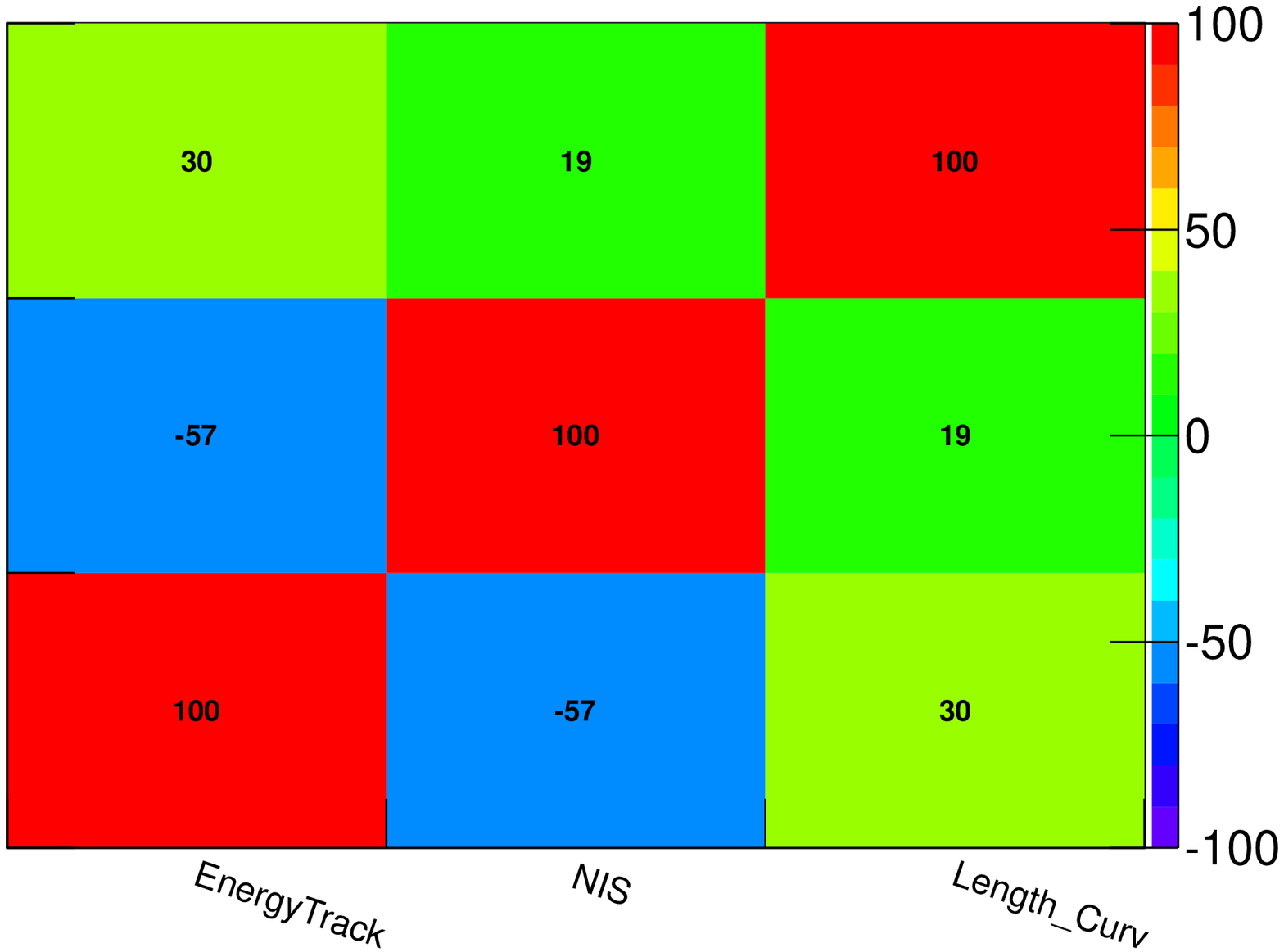}
\includegraphics[scale=0.38,angle=0]{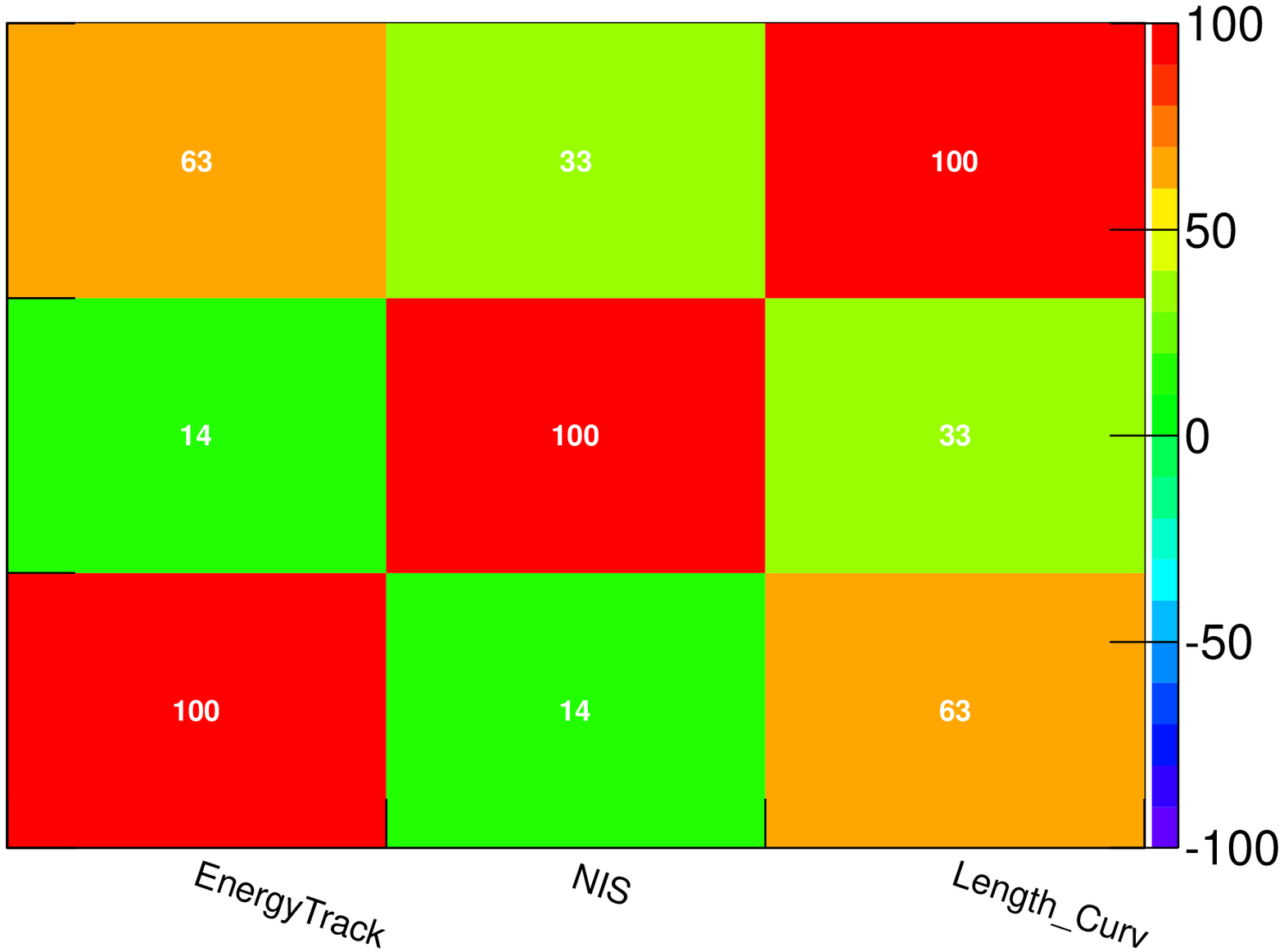}
\caption{Correlation matrix of the different observables, for fluorine recoils (left) and electron recoils (right). 
The various observables are indicated on the plot : $E_{ion}$, NIS and $L_c$.} 
\label{fig:CovMat}
\end{center}
\end{figure}

\noindent{\bf Correlations}\\
Figure~\ref{fig:FullDiscri} (third row, second column)   presents the distributions in the 
(NIS, $L_c$) plane. This highlights the complementarity between these two observables. While the electron distribution extends to large curve
length values with a constant NIS value, fluorine recoils are characterized by short curve length, below 3 cm, and NIS value ranging 
between $10^{-3}$ and 1. To check for correlations, we present on figure \ref{fig:CovMat} the correlation matrix between the 
different observables, for fluorine recoils (left) and electron recoils (right). First, we notice that the ionization energy and the NIS
observable are not correlated for nuclear recoils (14\%) and strongly anticorrelated for electrons ($\sim -60\%$). 
On the contrary, the curve length is slightly correlated ($\sim 30\%$) for electrons and strongly for 
fluorine recoils ($\sim +60\%$). Eventually, the NIS and curve length are only 
slightly correlated ($\sim 30\%$ for fluorine recoils and $\sim 20\%$ for electrons). Moreover, their correlation with the energy are of
opposite sign and clearly different for electron and fluorine recoils. Hence, the use of these discriminant observables seem to be
promising for e/recoil discrimination. From this discriminant study, we can conclude that the event tagging will be very efficient above
$\sim$ 20 keV, even with a straightforward cut-based analysis (sec.~\ref{sec:oldschool}). Below 20 keV, an advanced data analysis is needed
and we propose a boosted decision tree method for this purpose (sec.~\ref{sec:newschool}).

\begin{figure}[h]
\begin{center}
\includegraphics[scale=0.38,angle=0]{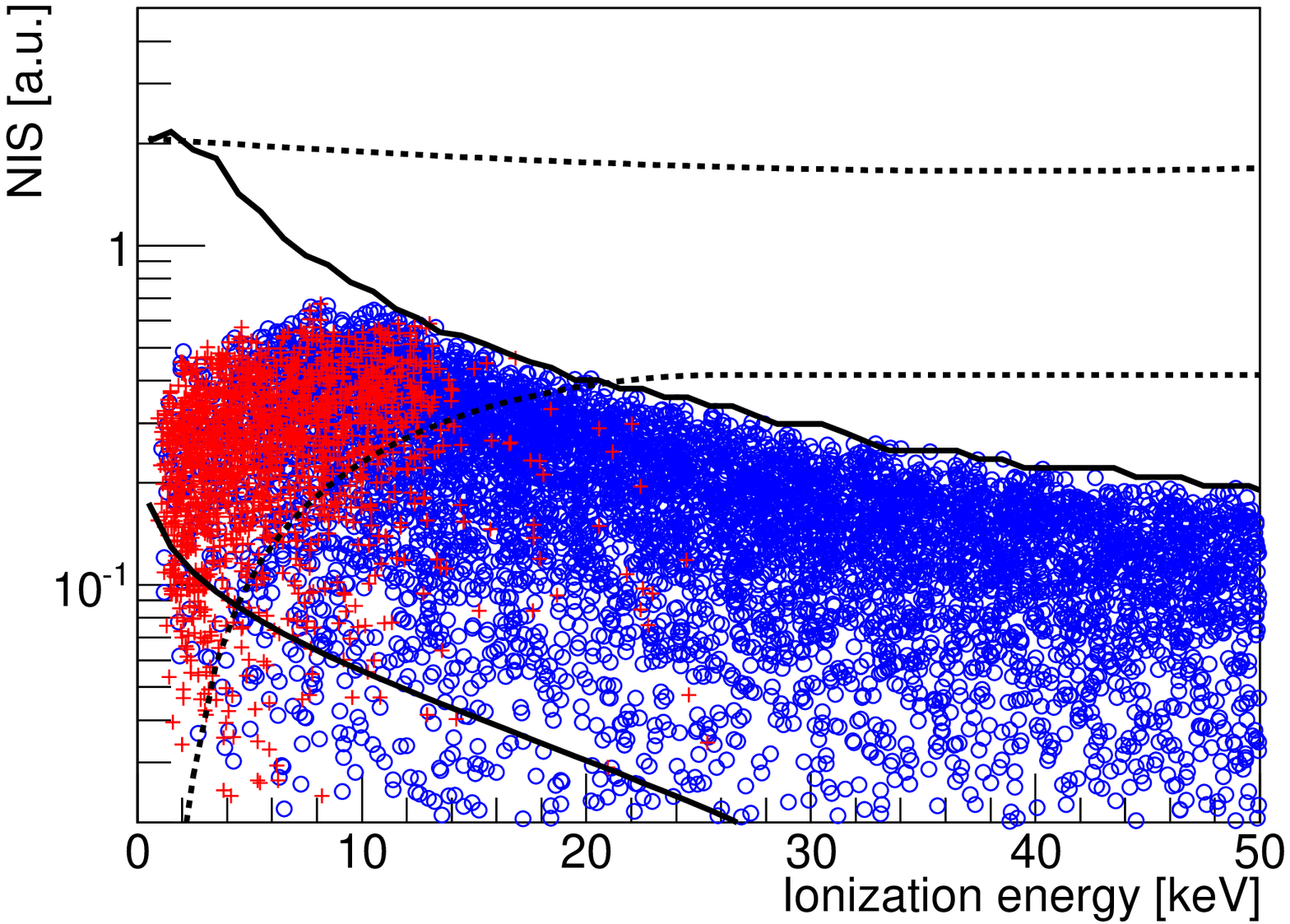}
\includegraphics[scale=0.38,angle=0]{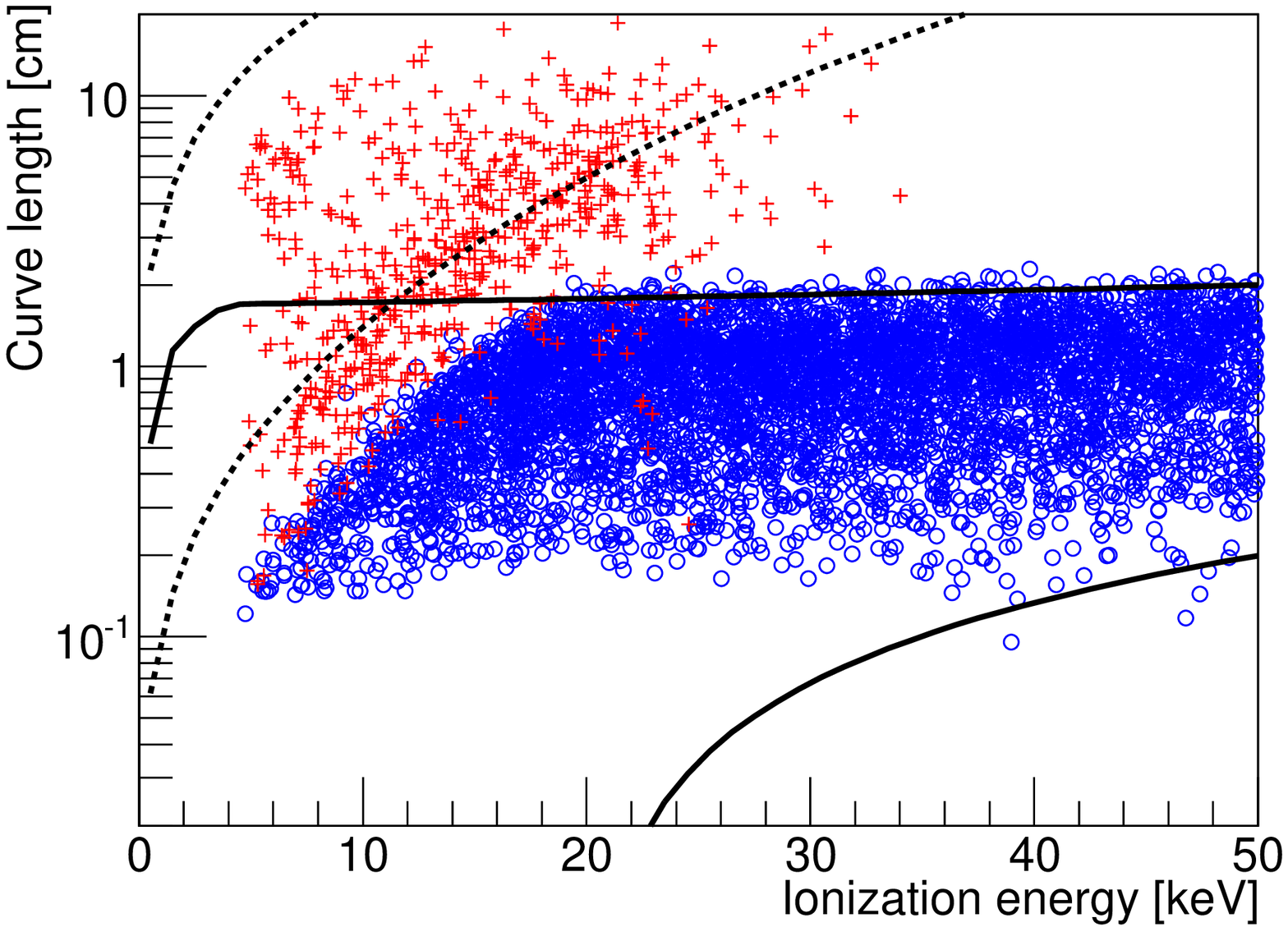}
\includegraphics[scale=0.38,angle=0]{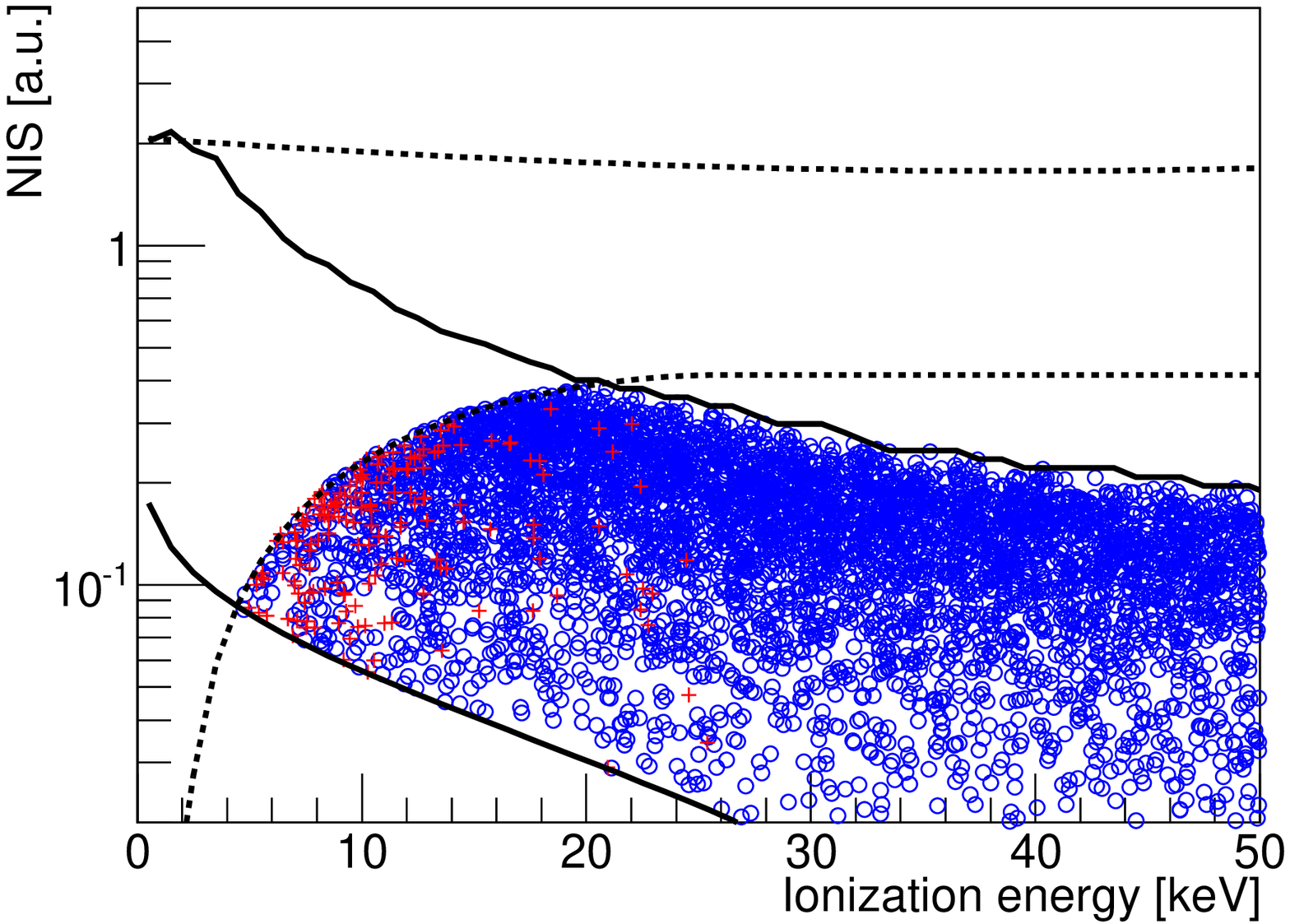}
\includegraphics[scale=0.38,angle=0]{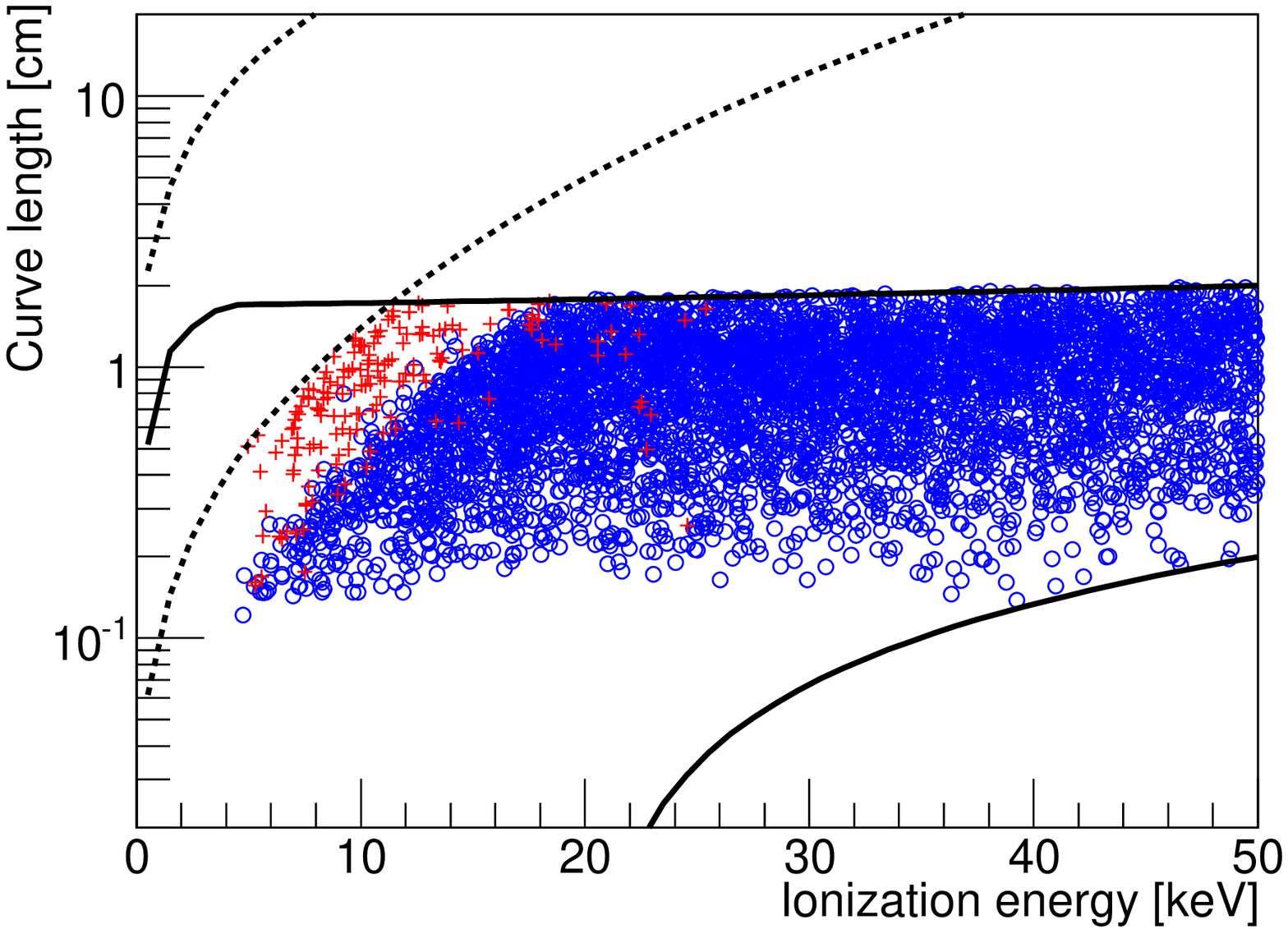}
\caption{Event distributions when selecting events within the 95\% CL recoil acceptance region on various  observables. 
Blue points present the events simulated as nuclear recoil whereas the red crosses present the simulated electrons. 
Left upper panel : (NIS, ionization energy) plane for a selection on the curve length observable only.  
Right upper panel :  (curve length, ionization energy)  for a selection on the NIS observable only.  
Lower panels : (NIS, ionization energy) plane and (curve length, ionization energy) plane after a selection on both NIS and curve length
observables.} 
\label{fig:HardCutNISetRange}
\end{center}
\end{figure}

\section{Sequential analysis}
\label{sec:oldschool}
The discriminant observables ($E_{ion}$, $L_c$ and NIS) are first used within the framework of a sequential analysis. The
goal is to evaluate the signal efficiency ($\mathscr{A}$), {\it i.e.} the fraction of nuclear recoils accepted by the analysis, as well as the
background rejection power (R). In the following, the rejection power R reads as :
\begin{equation}
R (E_{ion})= \frac{N_{e}}{N_{e}^{selec}}
\end{equation}
where $N_{e}$ is the number of electrons in the data sample, of a given energy for instance, and $N_{e}^{selec}$ after applying the
electron/recoil discrimination. Note that the rejection power should be defined, for a given confidence level, with an upper value on 
$N_{e}^{selec}$, following a Poisson statistics. We nevertherless stick to this standard definition for the sake of comparison with
rejection powers from others Dark Matter experiments.

\subsection{Defining cuts on the observables}
To define the cuts on the observables, the 95\% CL region on each observable ($L_c$ and NIS) have been evaluated as a function of the ionization energy. The
sequential analysis consists in rejecting all event outside the nuclear recoil acceptance region, defined as the region within the 
95\% CL nuclear recoil region minus its intersection with the 95\% CL electron region.\\
As an illustration, figure~\ref{fig:HardCutNISetRange} presents event 
distributions when selecting events within the 95\% CL recoil acceptance region on various  observables. 
Blue points present the events simulated as nuclear recoil whereas the red crosses present the simulated electrons. 
The left upper panel presents the event distribution 
in the (NIS, ionization energy) plane, when selecting events within the 95\% CL recoil acceptance region on the curve length observable 
only.  It can first be noticed that a substantial electron contamination remains within the recoil acceptance region. However they lie within the electron acceptance region and a large fraction will be removed 
by a selection on the NIS variable. Figure~\ref{fig:HardCutNISetRange} (upper right panel) presents the event distribution 
in the (curve length, ionization energy) plane, when selecting events within the 95\% CL recoil acceptance region on the NIS observable 
only.  Most electrons accepted by the NIS selection will be rejected by the curve length cut as they lie 
outside the recoil acceptance region, while loosing very few nuclear recoils. This highlights the complementarity between these two
discriminant observables. It is worth emphasizing that the NIS cut implies a high energy threshold ($\sim$ 4 keV), 
much larger than in the case of $L_c$ cut ($\lesssim 1 \ {\rm keV}$). The efficiency at low energy is also expected to be rather small in this case.\\
As a matter of fact, the e/recoil discrimination must include a selection on both discriminant observables (NIS and $L_c$). The recoil
acceptance is now defined in a 3D observable space (NIS, $L_c$ and $E_{ion}$), with cuts an the projection on each plane as
discussed above. Figure~\ref{fig:HardCutNISetRange} (lower panels)
presents the result of this selection in the (NIS, ionization energy) plane and in the ($L_c$, ionization energy) one. It can be seen that
the combined selection allows to reduce the electron contamination within the recoil acceptance region. However, the 4 keV threshold
associated to the NIS selection remains, as only hard-cuts are used in this analysis.

\begin{figure}[t]
\begin{center}
\includegraphics[scale=0.38,angle=0]{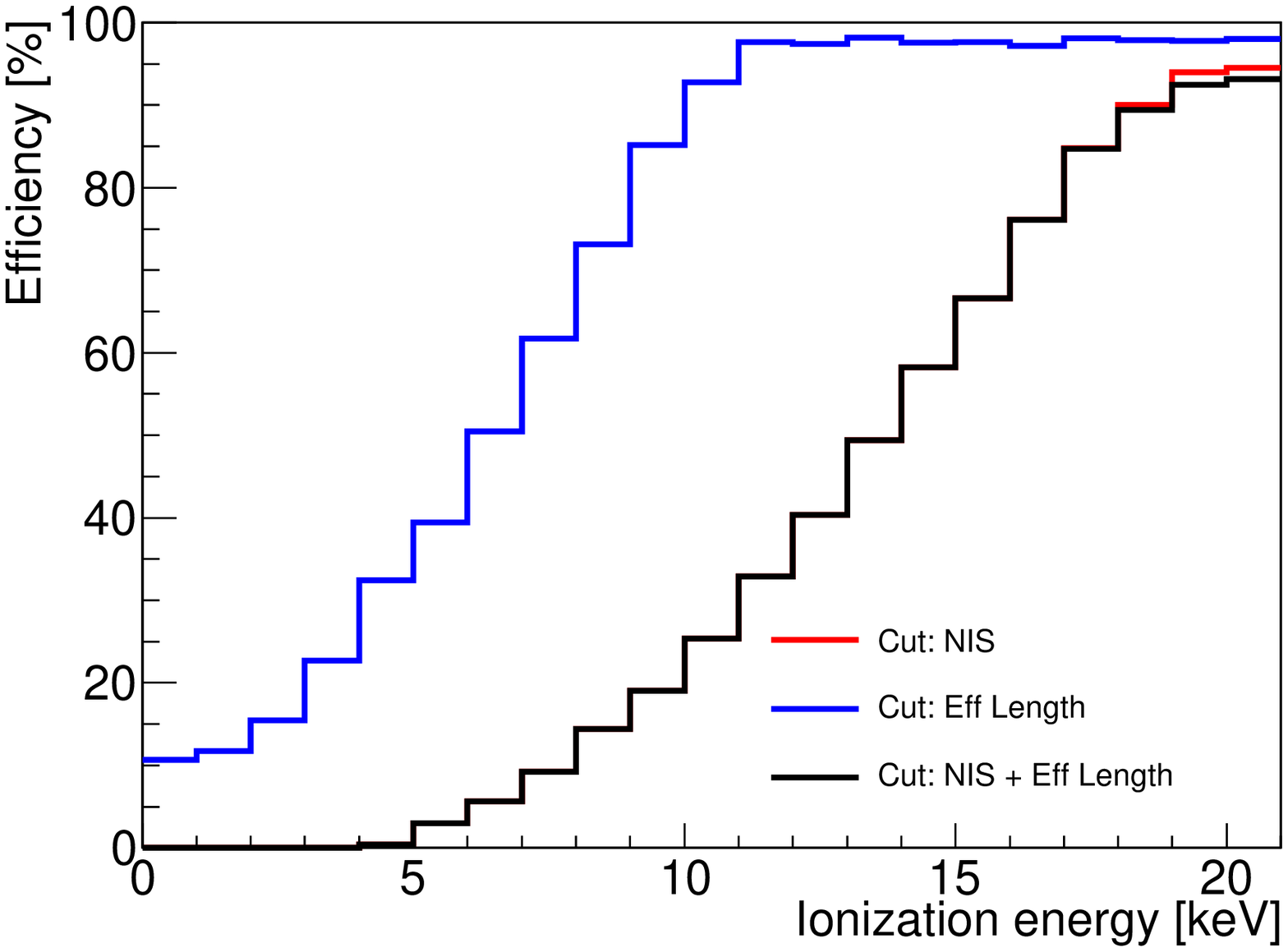}
\includegraphics[scale=0.38,angle=0]{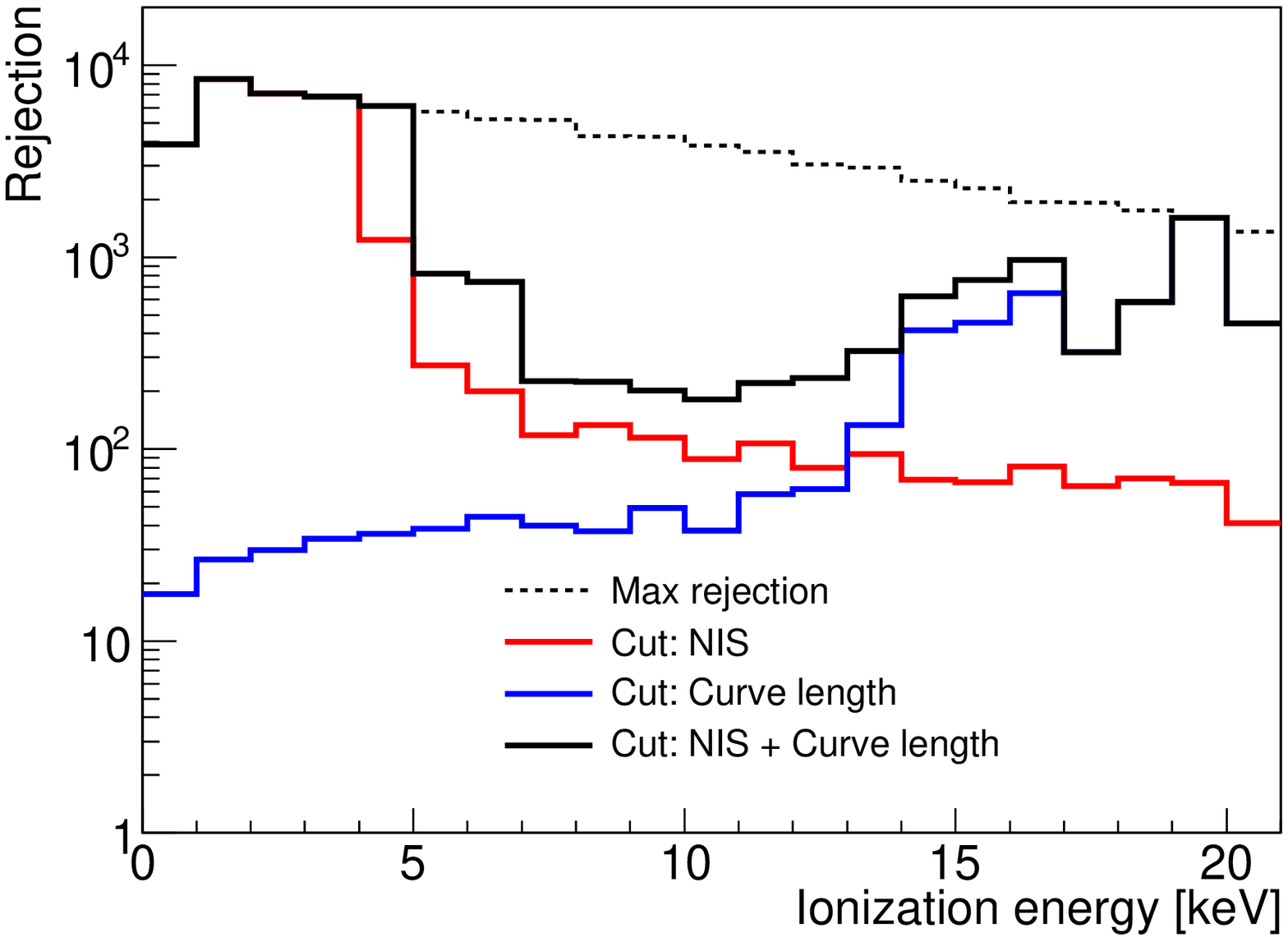}
\caption{Efficiency (left panel) and the rejection power (right panel) as a function of the ionization energy,   
in the case of a cut on the curve length only (blue),  
on the NIS observable only (red) and for a combined cut on the two observables 
(black).} 
\label{fig:HardCutNISetRangeEff}
\end{center}
\end{figure}

\subsection{Rejection and efficiencies}
Figure~\ref{fig:HardCutNISetRangeEff} presents the efficiency (left panel) and the rejection power (right panel) as a function of the ionization energy. 
In the case of a cut on the curve length only (fig.~\ref{fig:HardCutNISetRangeEff} blue curves), the efficiency raises from 10\% to 95\% when the ionization energy increases from 0 to 10 keV. The rejection power is higher than 100 above 15 keV.
For a cut on the NIS observable only (fig.~\ref{fig:HardCutNISetRangeEff} red curves), it is worth emphasizing that a 4 keV threshold is
observed, above which the efficiency increases with the ionization energy to reach 95\% at 20 keV. 
Due to this energy threshold, the rejection values below 4 keV are not relevant.  For a combined cut on the two observables 
(fig.~\ref{fig:HardCutNISetRangeEff} black curves), the efficiency is very close to the one of the NIS cut, as it is the most stringent 
cut at low energy. In particular, the 4 keV threshold stands as a major drawback of this analysis strategy. Between 4 and   14 keV, the rejection power is
dominated by the effect of the NIS cut, in the vicinity of 100. Above 14 keV, the effect of the $L_c$ cut becomes dominant and the
rejection power of the combined cut is increasing, reaching 1000 at 20 keV.  Above this energy, the e/recoil is expected to increase with
energy as the two distributions start to be well separated (see fig.~\ref{fig:LVsLRaw}).

\section{Event tagging  by Boosted Decision Tree analysis}
\label{sec:newschool}
Above 20 keV (ionization), a sequential analysis is probably sufficient to achieve high rejection power against electron together with a high recoil
efficiency. Indeed, the two distributions are well separated in the observable space (NIS, $L_c$ and $E_{ion}$). However, at low
energy, the efficiency obtained with such an analysis starts decreasing with decreasing energy. At 15 keV (ionization), it is of the order
of 50\% and a 4 keV threshold is observed. For rare event search, such as WIMP dark matter, we argue that this should be improved. In
particular, other effects are expected to raise the energy threshold ({\it e.g.} quenching factor, 3D reconstruction, ...) and the e/recoil
discrimination should be optimized to limit its effect at low energy.\\
For this purpose a multivariate analysis is proposed in order to improve electron/recoil discrimination at low energy. It is a Boosted
Decision Tree (BDT) analysis, based on the same discriminant observables.

\subsection{Introduction to BDT analysis}
Boosted Decision Tree (BDT) analysis belongs to machine learning techniques and is widely used to treat data 
in high energy physics ({\it e.g.} \cite{miniboone,d0.top,Conrad:2006ip}). It is an extension of the commonly used  cut-based selection strategy into a multivariate technique. 
Decision Tree analysis can be seen as a data classifier and is often used for signal/background discrimination. 
Indeed, as most events do not have all characteristics of either signal or background, the principle  of Decision Tree is 
to keep events that fail a given criterion and check for other observable discriminants. Trees can then be boosted to combine weak
classifiers into a new one with smaller error rate \cite{bdt}.\\
The result of a BDT analysis is given by a forest of $N_{trees}$ decision trees ($T_k$) combined in a unique variable $X_{BDT}$, 
containing all the information. It reads as : 
\begin{equation}
X_{BDT} = \sum_{k=1}^{N_{trees}} \alpha_k T_k(\tilde{O}^l)
\end{equation}
where $\tilde{O}^l$ are the discriminant observables and $\alpha_k$ the weight of each tree $T_k$.\\
Boosting is then applied to increase the weight of misclassified events in order to reduce 
the misclassification rate of the next tree. Note that the misclassification rate can be reduced to zero, but at the price of overtraining. This appears when trees contain to many leafs, 
with a low number of events. In this case, the algorithm is sensitive to statiscal fluctuations of the training sample. A Kolmogorov
statistical test on cumulative distributions associated to the test sample and to the training sample is used to check 
for overtraining.\\
In a previous paper \cite{billard.track}, we have proposed a new method to optimize the sense recognition efficiency for 
3D track reconstruction of directional data. We extend this idea to improve the recoil efficiency of e/recoil discrimination 
at low energy. In the following, we acknowledge the use of the TMVA software  \cite{tmva}.

\subsection{BDT analysis for e/recoil discrimination}
To apply BDT analysis to e/recoil discrimination, we use the same
observables as for the sequential analysis, namely NIS, $L_c$ and $E_{ion}$. 
Two data populations are used (electrons and nuclear recoils). Trees have been trained with  10$^5$ events (nuclear recoils and electrons). 
To avoid overtraining while maximizing the performance of the BDT analysis, 
a minimum number of 100 events per leaf is required as well as a maximum number of tree levels of 50. We found this configuration to be optimal as it stands at the limit of
overtraining. Indeed, following the Kolmogorov test statistic, we found a maximal distance between the train and test sample cumulative distributions of $5.87\times 10^{-3}$
corresponding to a p-value of $7.7\times 10^{-2}$. This result suggests a 2$\sigma$ agreement between the two distributions validating our statement that the Boosted Decision
Tree algorithm considered in the following is not overtrained.\\

\begin{figure}[t]
\begin{center}
\includegraphics[scale=0.38,angle=0]{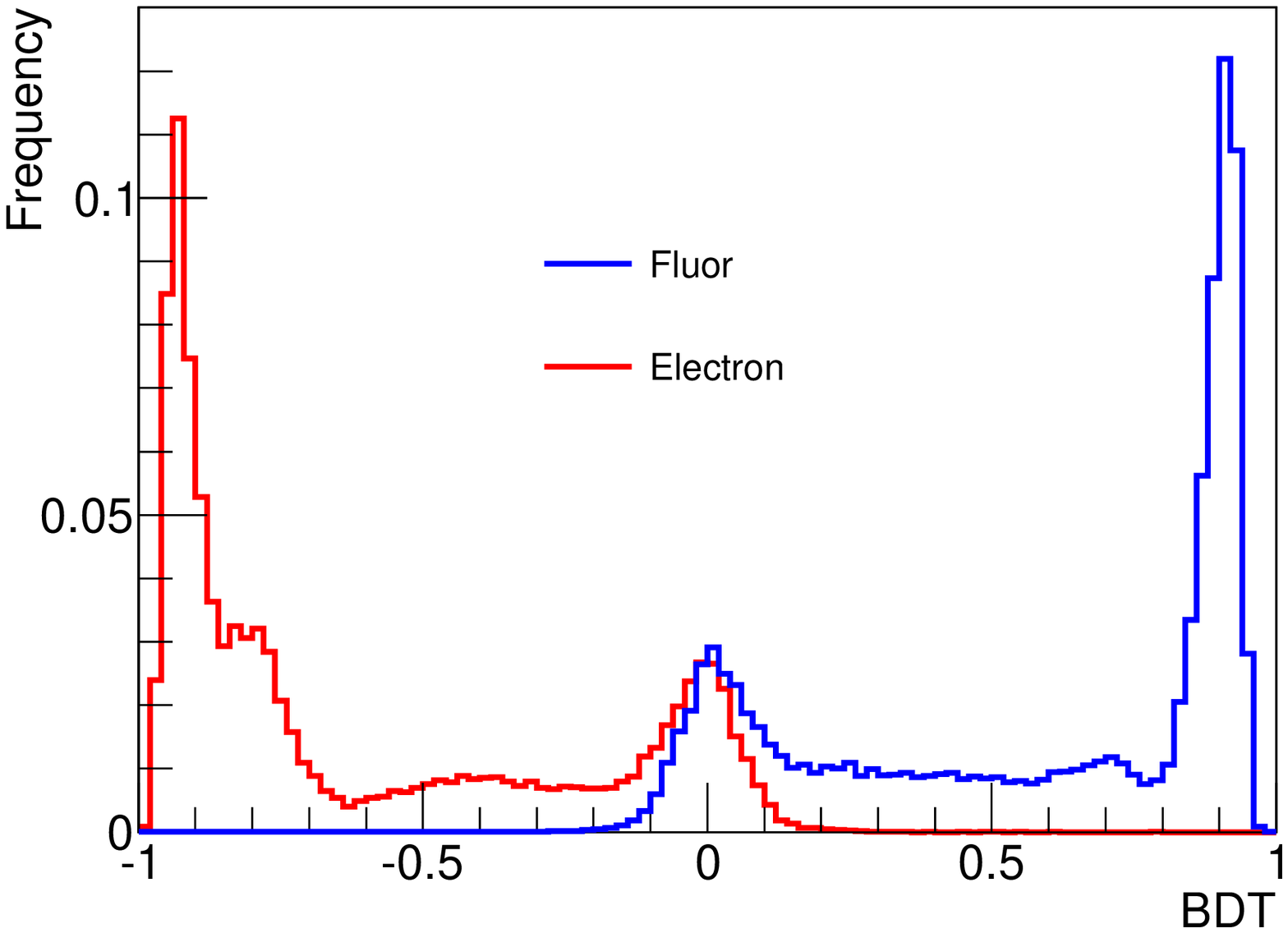}
\includegraphics[scale=0.38,angle=0]{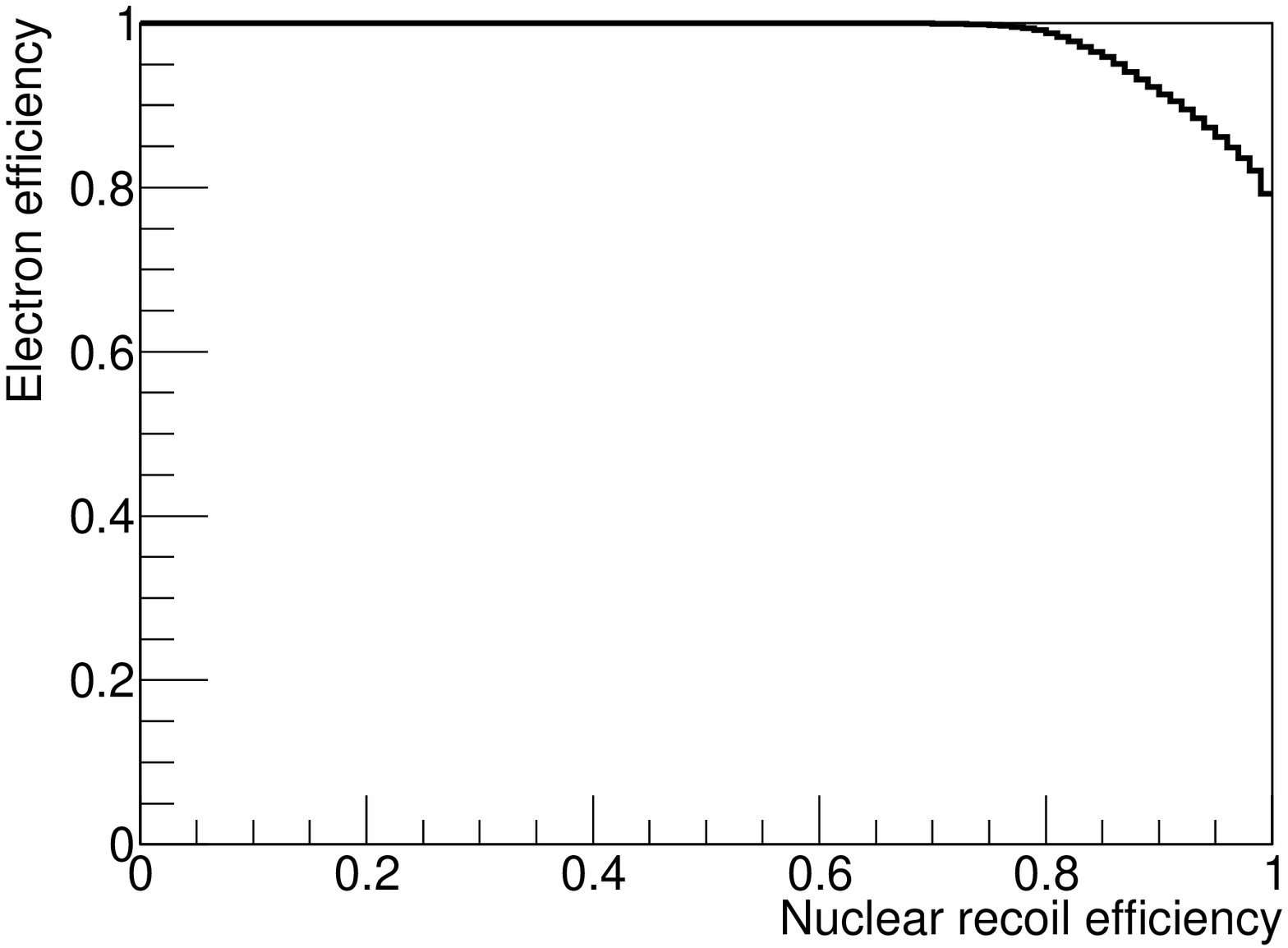}
\caption{Left : $X_{BDT}$ distributions for electrons (red) and nuclear recoils (blue). Right : 
electron efficiency as a function of the nuclear recoil efficiency in the case of this analysis applied on the simulated electron and nuclear recoil distributions.} 
\label{fig:BDTDistrib}
\end{center}
\end{figure}

Figure~\ref{fig:BDTDistrib} (left) presents the $X_{BDT}$ distributions for electrons (red) and nuclear recoils (blue). In the following, 
we aim at selecting nuclear recoils while rejecting electrons. Hence events need to be tagged either as electrons or nuclear recoils. 
Such  event tagging is achieved by applying a cut on the value of $X_{BDT}$. This implies, for each cut value,  a given efficiency and
rejection that can then be easily estimated. Figure~\ref{fig:BDTDistrib} (right) presents the electron efficiency as a function of the nuclear recoil efficiency in the
case of this analysis is applied on the simulated electron and nuclear recoil distributions.
 We can conclude from this figure, that an electron efficiency above 90\% implies a nuclear recoil efficiency below 90\%.\\
There is two ways to use a BDT analysis :  either events tagged as electrons are rejected (sec.~\ref{sec:hard}) or all events are kept and
weighted according to their electron probability  (sec.~\ref{sec:prob}).

\begin{figure}[t]
\begin{center}
\includegraphics[scale=0.38,angle=0]{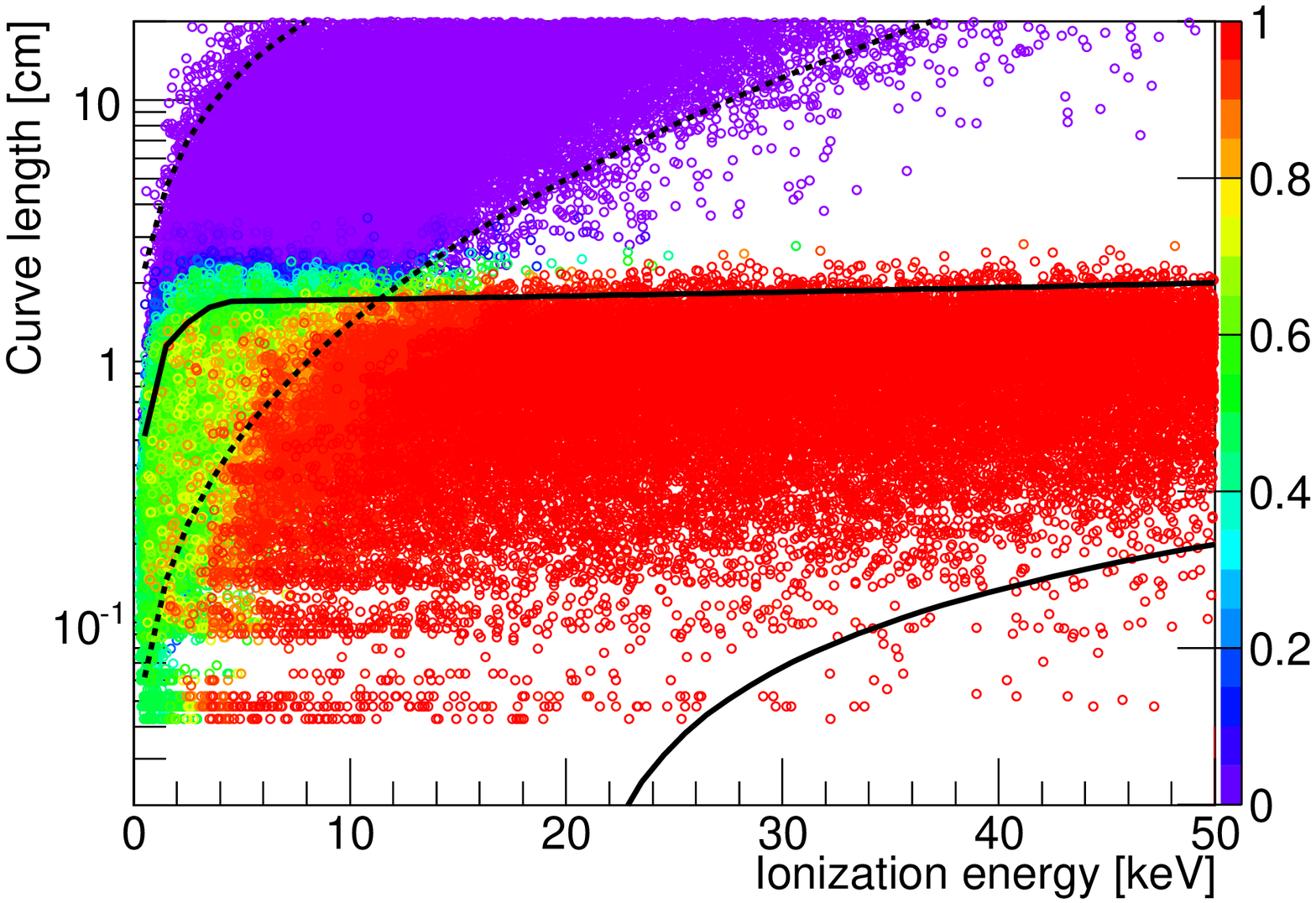}
\includegraphics[scale=0.38,angle=0]{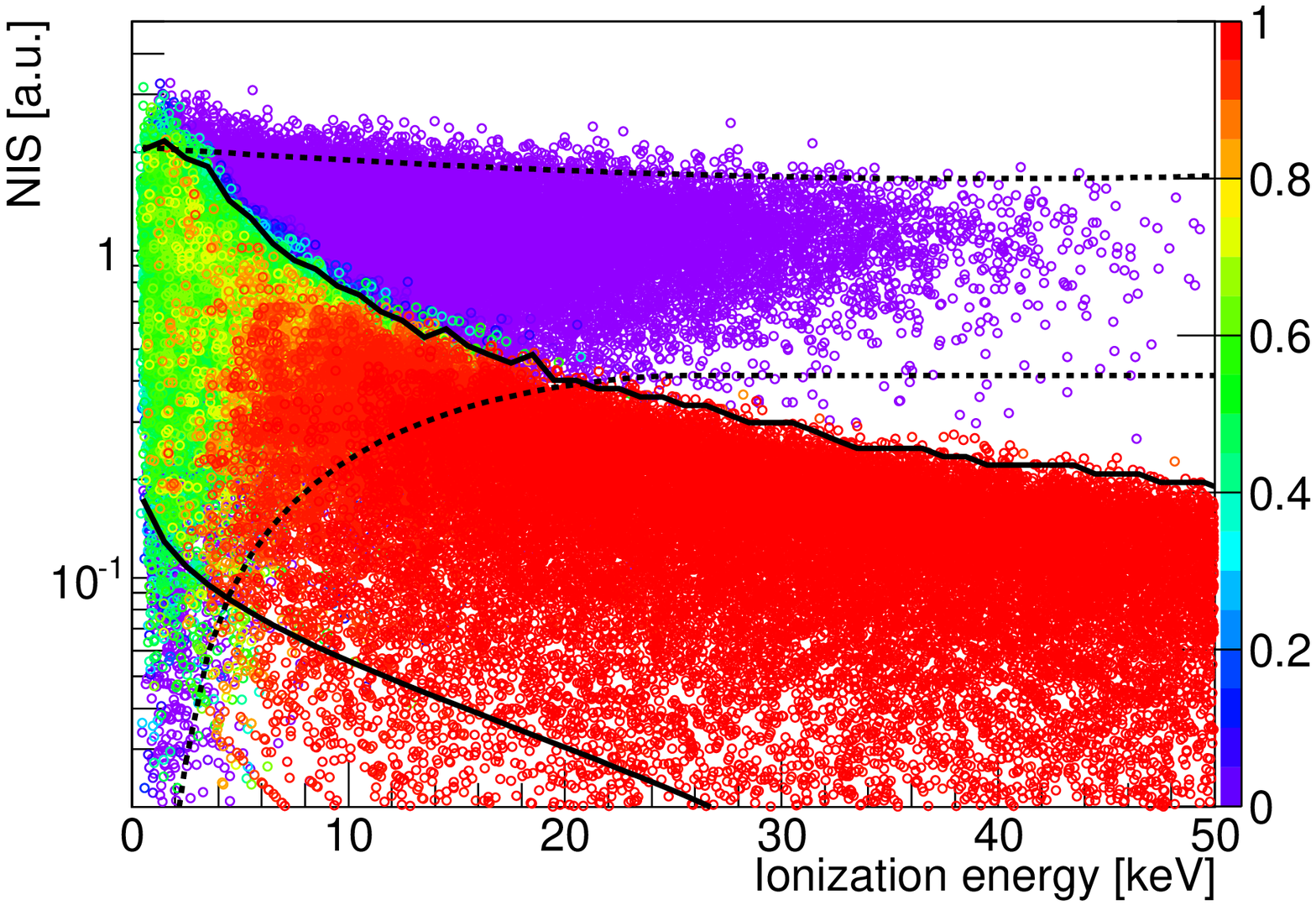}
\caption{Event distribution after event tagging : ($L_c$ vs $E_{ion}$) left (NIS vs $E_{ion}$) right. 
The color code indicates the value of $w_i$, {\it i.e.} the probability to be  nuclear recoil.} 
\label{fig:BDTProba}
\end{center}
\end{figure}

\subsection{Probability weighting method}
\label{sec:prob}
For a given event, the value of $X_{BDT}$ is related to its probability to be a nuclear recoil $w_i$, via the Bayes' theorem. It is
 defined from the conditional distributions $f(X_{BDT}|{\rm electron}) \equiv f_e(X_{BDT})$ and $f(X_{BDT}|{\rm fluor}) \equiv f_f(X_{BDT}) $ 
 (cf. figure~\ref{fig:BDTDistrib}) as:
\begin{equation}
w_i = \frac{f_f(X^i_{BDT}) }{f_e(X^i_{BDT}) + f_f(X^i_{BDT})}.
\end{equation}
Figure~\ref{fig:BDTProba} presents all events in the 
(NIS, $E_{ion}$) plane (right) and in the  ($L_c$, $E_{ion}$) one (left) where the color code is related to their probability to be a nuclear recoil $w_i$. 
 Acceptance regions from the
sequential analysis are recalled to ease comparison. Most events in the  95\% acceptance  nuclear recoil region are tagged as such. 
Above 20 keV, the discrimination is almost perfect and events are tagged either as electrons or nuclear recoils with a null
misidentification. Interestingly, below 20 keV a fraction of events are identified as nuclear recoils whereas  they lie well outside the
nuclear recoil acceptance region. Hence, they would have been missed by a sequential analysis. Below 10 keV a large numbers of events are
associated with a probability  $0 < w_i <1$ associated with an ambiguous identification. Whereas a hard cut on the value of BDT would
remove them and hence decrease the efficiency, it is possible to decrease their effect on Dark Matter search, by taking into account their
probability to be a nuclear recoil or an electronic recoil.\\
\begin{figure}[t]
\begin{center}
\includegraphics[scale=0.5,angle=0]{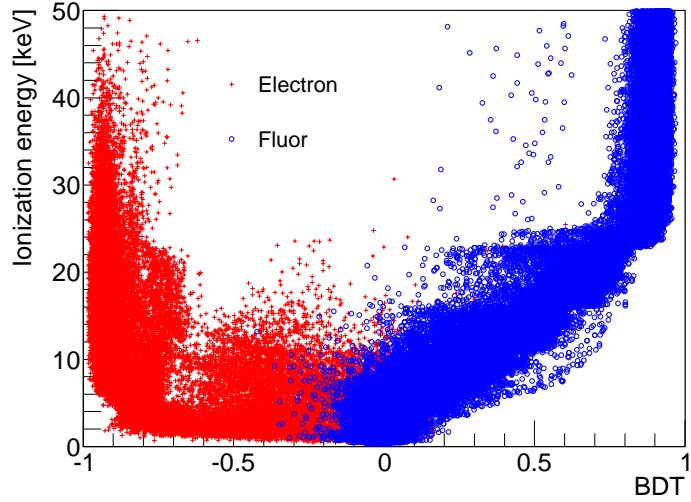}
\caption{Distribution in the (BDT, $E_{ion}$) plane for electronic events (red crosses) and fluorine recoils (blue circles).} 
\label{fig:BDTVsEnergy}
\end{center}
\end{figure}
To perform such  analysis without using hard cuts on the $X_{BDT}$ value, we defined the following likelihood function aiming at fitting all the recorded events in 
the $\{ E_r (\equiv E_{ion}), \Omega_r, L_c, NIS\}$ parameter space. The observables $E_{ion}$, $L_c$ and $NIS$ can be transformed into a unique variable, $X_{BDT}$, 
following the distributions shown on figure~\ref{fig:BDTDistrib} (left). However, we found that there is a strong correlation between $X_{BDT}$ and $E_{ion}$ 
(see fig.~\ref{fig:BDTVsEnergy}) implying that
the likelihood function from \cite{billard.ident,billard.profile,billard.exclusion} must be modified as follows :
\begin{eqnarray}
\mathscr{L}(\vec{\theta})    = &  \frac{e^{-\mu_{\rm tot}}}{N!} \\ \nonumber
 & \times \prod_{i=1}^{N} 
 \left[\left(\mu_sS(\vec{R}_i)+\mu_nN(\vec{R}_i)\right)\times f_f(X^i_{BDT}|E_i)  +  \mu_eE(\vec{R}_i)\times f_e(X^i_{BDT}|E_i)\right]
\end{eqnarray}
with $\mu_{\rm tot}$ defined as : 
\begin{equation}
\mu_{\rm tot} = \mu_s + \mu_n +\mu_e 
\end{equation}
where $\mu_s$, $\mu_n$ et $\mu_e$ correspond to the expected mean number of WIMP, neutron and electron events respectively, which are part
of the free parameters ($\vec{\theta}$) of the likelihood function and $N$ is the total number of events measured.\\
The function $S(\vec{R}_n)$, $N(\vec{R}_n)$ and $E(\vec{R}_n)$
correspond to the $d^2N/dE_rd\Omega_r$  distributions associated with WIMP, neutron and electron events respectively. 
$\vec{R}_i$ refers to the characteritics of the $i$-th event (recoil energy $E_r$ and direction $\Omega_r$). 
The functions $f_f(X_{BDT}|E)$ and $f_e(X_{BDT}|E)$ correspond to the
normalized conditional probability distributions of $X_{BDT}$ depending on the energy. These distributions can easily be estimated from the figure~\ref{fig:BDTVsEnergy} which
shows that there is a strong correlation between $X_{BDT}$ and $E_{ion}$. Indeed, greater is the ionization energy, better is the discrimination between the two popullations
leading to  distributions more peaked around -1 for the electrons and +1 for the fluorine recoils. The non-correlation between $X_{BDT}$ and $\Omega_r$ is simply
explained by the fact that $L_c$ and $NIS$ are weakly correlated with the orientation of the track in the detector frame due to primary electron diffusion and detector
limitations (pixel size and time sampling of the anode). Moreover, from the detector frame to the galactic frame, there is several rotations smearing out 
any residual correlations between the direction of the recoil in galactic coordinates and $X_{BDT}$.\\

Hence, this analysis strategy allows to account for all events which has the benefit of maximizing the efficiency and then the Dark Matter reach of a given 
directional detector (see sec.~\ref{sec:con}).
 This way, if no events is observed, we keep a 100\% efficiency. Estimation of the effect of such analysis on the discovery and exclusion potential
 is beyond the scope of this paper.

\subsection{Cut-based BDT method}
\label{sec:hard}
Another way to use BDT analysis is to select only events above a given value of $X_{BDT}$. Hereafter, this  event tagging method is referred to as cut-based BDT analysis. 
Efficiencies and rejections are then estimated as explained above (see fig.~\ref{fig:BDTDistrib}). All events associated 
with $X_{BDT}$ value greater than a threshold value, X$_{\rm cut}$, are tagged as
nuclear recoils whereas all events with BDT value below this value are rejected.

\begin{figure}[t]
\begin{center}
\includegraphics[scale=0.38,angle=0]{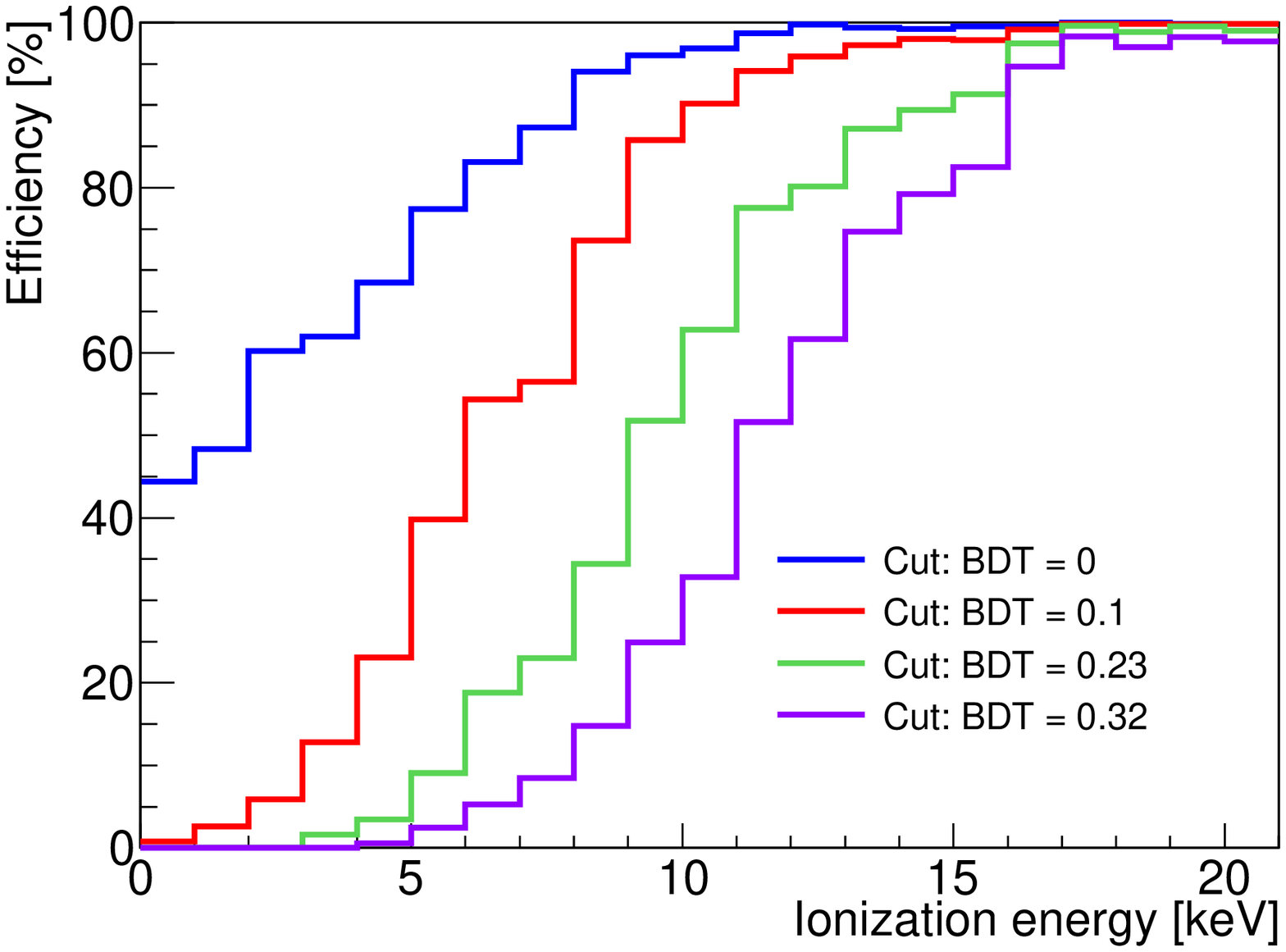}
\includegraphics[scale=0.38,angle=0]{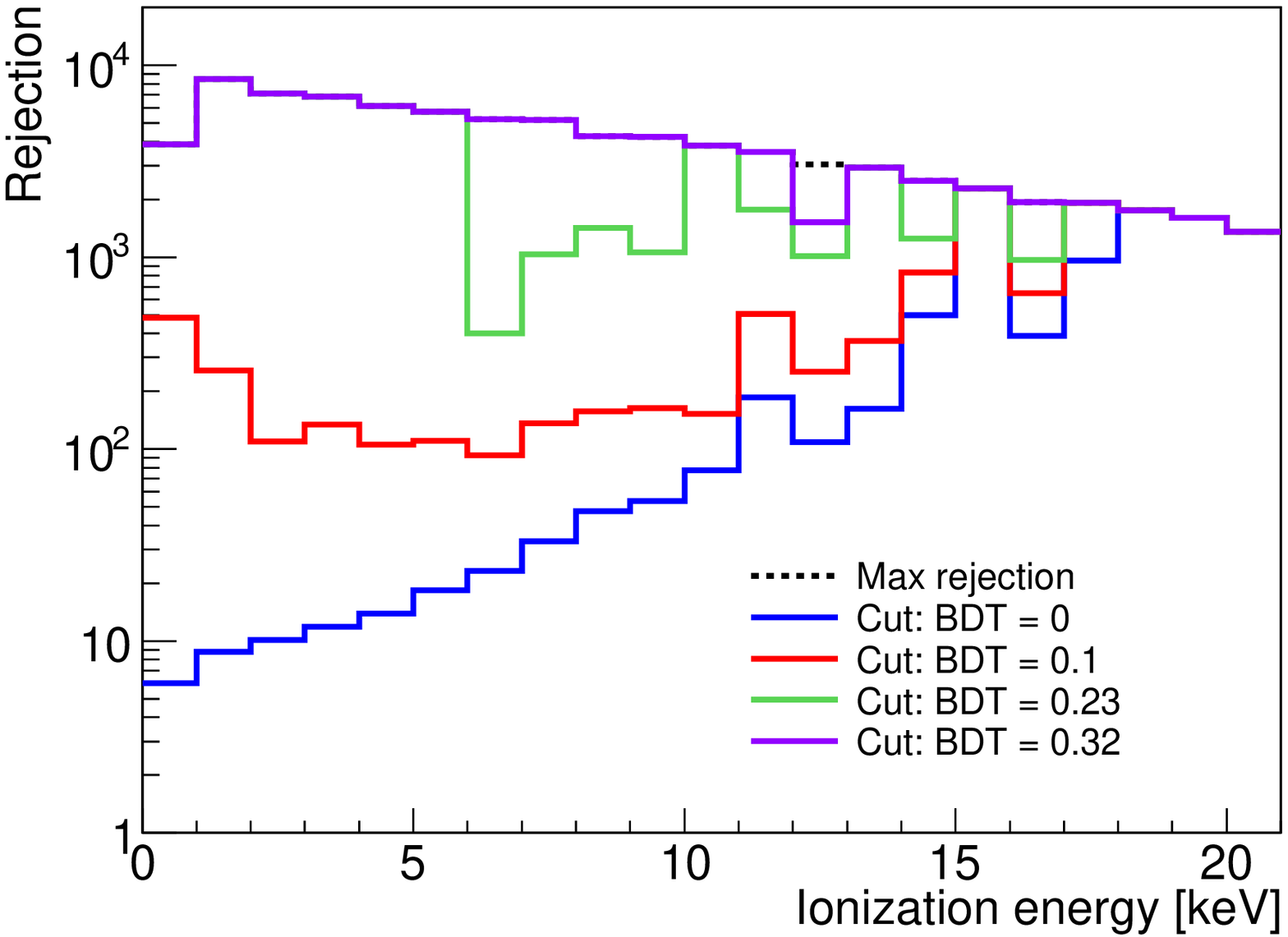}
\caption{Efficiency (left) and the rejection (right) as a function of the ionization energy, for several
values of X$_{\rm cut}$ :  X$_{\rm cut}$ = 0 (blue), X$_{\rm cut}$ = 0.1 (red), X$_{\rm cut}$ = 0.23 (green) et X$_{\rm cut}$ =
0.32 (purple).} 
\label{fig:BDTCut0Eff}
\end{center}
\end{figure}

Figure~\ref{fig:BDTCut0Eff} presents the efficiency (left) and the rejection (right) as a function of the ionization energy, for several
values of X$_{\rm cut}$. For a value  X$_{\rm cut}$ = 0, we notice that there is no threshold effect induced by this discrimination
strategy. At 0 keV, the efficiency is $\sim 50 \%$ and increases with ionization energy to reach 90\% around 10 keV. 
For this selection, the rejection against electrons starts at 10 at 0 keV and reaches 1000 around 20
keV. For a value  X$_{\rm cut}$ = 0.1, the efficiency reaches 50\% at 5 keV (ionization) and implies a rejection against electrons always 
above 100. As expected, increasing the value of X$_{\rm cut}$ results in a loss of efficiency and an increase of rejection. We caution that the higher value of rejection (X$_{\rm cut}$=0.32) are limited by the number of events 
generated in this simulation. They must hence be taken as a minimum value.\\ 
Interestingly, the BDT analysis alleviates the threshold effect observed in fig.~\ref{fig:HardCutNISetRangeEff} and allows for high
rejection power together with high efficiencies below 20 keV (ionization). It is worth emphasizing this upper value chosen for this
analysis. First, and as  stated above, the discrimination above 20 keV (ionization) efficient even with a sequential analysis. 
The expected rejection values are very high ($> 10^{3-4}$), noticing that these values correspond to a flat electron spectrum, between 0 and 50 keV. Moreover, 20 keV (ionization) corresponds to about 50 keV (recoil) when taking
into account the Fluorine Ionization Quenching Factor.\\
We conclude that cut-based BDT analysis provides a powerful way to discriminate electron from nuclear recoils at low energy. This event
tagging strategy is of general interest for Dark Matter search and could be
adapted for various e/recoil discrimination strategies. {\it In fine}, the level of
rejection needed can be  tuned by choosing  the value of  X$_{\rm cut}$. It is then a matter of choice   to set a priority 
either on the efficiency or on the rejection. This issue is closely related to radiopurity of the chosen material and Dark matter search strategy. Indeed, as outlined in  \cite{billard.profile,billard.exclusion},  directional detection could accommodate to a sizeable 
background contamination, thanks to the  evaluation of the double-differential spectrum $\mathrm{d}^2R/\mathrm{d}E_R\mathrm{d}\Omega_R$. 
Moreover, as shown in \cite{billard.track}, the reconstruction of the localization of the vertex of event 
in the detector volume is a key issue in  order to discriminate surface events, mostly coming from the radioactivity 
from the surrounding detector material. It allows to define a fiducial cut which 
will considerably reduce the background event contamination of the data, prior to the e/recoil discrimination.

\section{Consequences for Dark Matter search}
\label{sec:con}
A  cut-based BDT analysis would provide a high rejection against electrons and a good efficiency, depending on the choice of cut on the BDT
variable. In this section, we investigate the consequences for Dark Matter search potential in the case of the 30 kg.year 
MIMAC detector.  First, we evaluate the event rate in the ionization energy range, taking into account the selection efficiency from this e/recoil
discrimination. In the following we note the latter as $\mathscr{A}(E_{\rm ion})$. It depends both on the ionization energy 
(fig.~\ref{fig:BDTCut0Eff}) 
and on the value of
BDT$_{cut}$.  The total number of WIMP events, $\mu_s$, reads as :  
 \begin{equation}
 \mu_s = \int_{E_{\rm ion}^1}^{E_{\rm ion}^2}\left[\int_{0}^{\infty}\frac{dR}{dE_{\rm ion}}(E_{\rm ion}')\times K(E_{\rm ion}',E_{\rm ion})\times\epsilon(E_{\rm ion}') dE_{\rm
 ion}'\right]\times
 \mathscr{A}(E_{\rm ion})dE_{\rm ion}
 \end{equation}
where $\epsilon$ is the detection efficiency,  {\it i.e.} events having $N_c>1$ and fully enclosed within the detector,  
$Q(E_r)$ is the average value of the ionization quenching. $K(E_r,E_{\rm ion})$ is the energy resolution of the detector, taking into
account fluctuations in the energy loss channel, the number of electron/ion pairs and the amplification.\\
The selection efficiency, $\mathscr{A}(E_{\rm ion})$,  is evaluated in the following within the framework of a sequential analysis 
(sec.~\ref{sec:oldschool}), as well as a boosted decision tree method (sec.~\ref{sec:newschool}). $\mu_s$ is Monte Carlo evaluated with  
10$^5$   WIMP events with an energy range between  0 and 160 keV.\\

 \begin{figure}[p]
\begin{center}
\includegraphics[scale=0.5,angle=0]{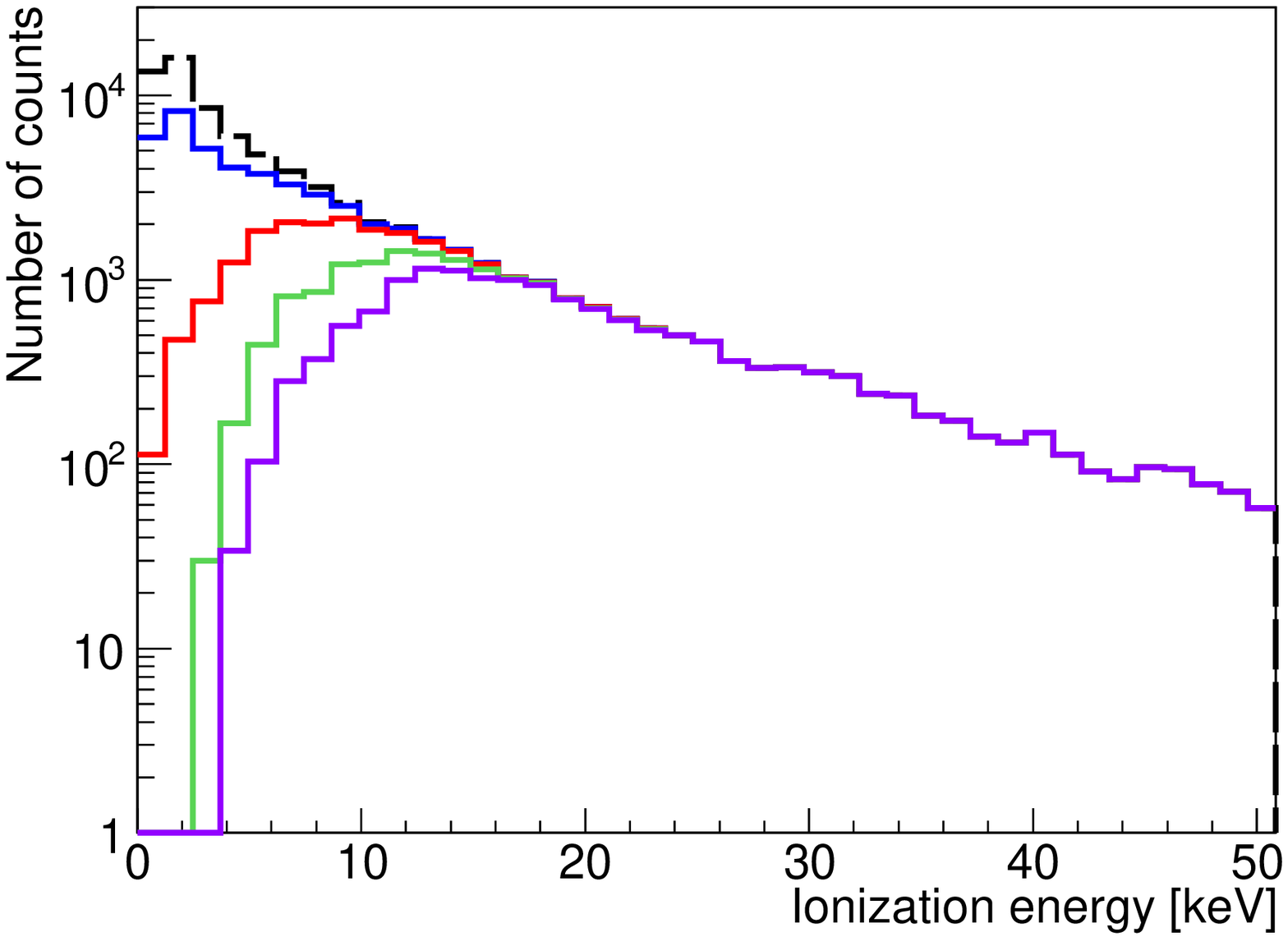}
\caption{Ionization energy spectra for WIMP mass 
$m_{\chi} = 100$ GeV.c$^{-2}$, for a rejection power : $R = 10$ (X$_{\rm cut}$ = 0) in blue, 
$R = 100$ (X$_{\rm cut}$ = 0.1) in red, $R = 1000$
 (X$_{\rm cut}$ = 0.23) in green et $R = 10000$ (X$_{\rm cut}$ = 0.32) in purple. The black distribution presents the case 
where no selection is applied on the BDT variable.} 
\label{fig:BDTCutSpectre}
\includegraphics[scale=0.5,angle=0]{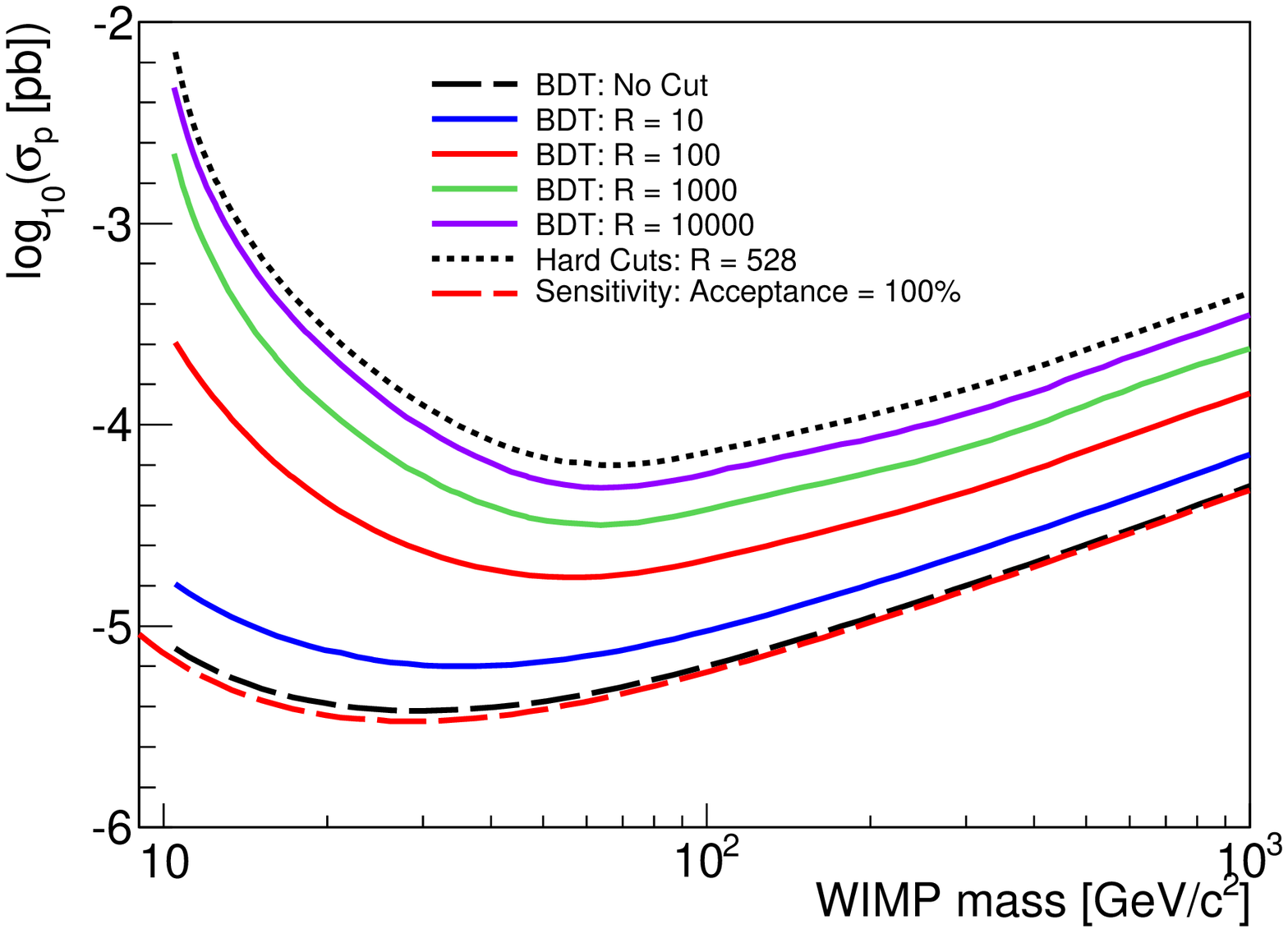}
\caption{Sensitivity of a 30 kg.year $CF_4$ directional detector (MIMAC). The result is presented as a
limit in ($\sigma_p,m_\chi$), where  $\sigma_p$ is the spin-dependent cross section on nucleon and $m_\chi$ the WIMP mass.  
The dashed red curve  represents the sensitivity without cuts.  The sensitivity for the loose cut case is presented 
as a dashed black curve. Rejections obtained with a BDT analysis are presented : 
$R = 10$ in blue, $R = 100$ in red,
  $R = 1000$ in green and $R = 10^4$ in purple. The sensitivity for sequential analysis, with a rejection R=528, is presented as a dotted
  black curve.} 
\label{fig:BDTCutPotentiel}
\end{center}
\end{figure}

Figure~\ref{fig:BDTCutSpectre} presents the ionization energy spectra for WIMP mass 
$m_{\chi} = 100$ GeV.c$^{-2}$, for a rejection power : $R = 10$ (X$_{\rm cut}$ = 0) in blue, 
$R = 100$ (X$_{\rm cut}$ = 0.1) in red, $R = 1000$
 (X$_{\rm cut}$ = 0.23) in green et $R = 10^4$ (X$_{\rm cut}$ = 0.32) in purple. The black distribution presents the case 
where no selection is applied on the BDT variable ($\mathscr{A}(E_{\rm ion}) = 1$). It shows the 
impact of : the loose selection (events having $N_c>1$ and fully enclosed within the detector), 
the ionisation quenching  factor (conversion from recoil energy to ionization energy) and the energy resolution. Note that would be the
case for the weighting probability method.\\
As expected, an increase of the rejection power, via a tighter cut on X$_{\rm cut}$, implies a loss of low energy events. For a rejection
$R = 10^4$ it corresponds to a 4 keV threshold. It is worth emphasizing that the rejection values cited hereabove correspond 
to a flat electron  spectrum, between 0 and 50 keV. Hence, these values must be taken as reference values and they must be treated with caution when compared to experimental values, obtained from a
$\gamma$-ray source \cite{Lopez:2011yv}.\\

Figure~\ref{fig:BDTCutPotentiel} presents the sensitivity of a 30 kg.year $CF_4$ directional detector (MIMAC) \cite{billard.exclusion}. 
The result is presented as a limit in ($\sigma_p,m_\chi$), where  $\sigma_p$ is 
the spin-dependent cross section on proton and $m_\chi$ the WIMP mass. The sensitivity is defined as the Poisonnian limit with 0 event
measured. The recoil energy range chosen for this sudy is 5-50 keV, which correspond to an ionization energy range of 0.5-17.5 keV, when
taking into account the ionization quenching factor as obtained from SRIM simulations \cite{srim}. The dashed red curve 
represents the sensitivity without cuts. It stands as the ultimate reach for such an exposure. 
The sensitivity for the loose cut case is presented as a dashed black curve. 
Interestingly, it is also the sensitivity expected for  the weighting probability method in the 0 event case.  
The effect of such a selection on the sensitivity of the MIMAC detector 
is shown to be weak. Rejections obtained with a BDT analysis are presented : 
$R = 10$ in blue, $R = 100$ in red,
  $R = 1000$ in green and $R = 10^4$ in purple. The sensitivity for sequential analysis, with a rejection R=528, is presented as a dotted
  black curve. It can be noticed that increasing the rejection results in a loss of sensitivity to low mass WIMPs. Reaching $R=10^4$, 
  the loss   of sensitivity is of the order of one order of magnitude at high WIMP mass. As stated above, the choice of the rejection power
  value, through the choice of the value of X$_{\rm cut}$, is related to the acceptable level of  background contamination.
Eventually, we highlight the fact that a Boosted Decision Tree e/recoil discrimination strategy would have 
 a rejection $\sim 20$ times higher than a sequential analysis, for the same Dark Matter exclusion reach.

\section{Conclusion}

Even though it could accommodate to a sizeable background contamination, electron/recoil discrimination 
remains a key and challenging issue as for direction-insensitive detectors. In the case of 
low pressure TPCs dedicated to directional search, the event background rate can  be reduced through fiducialization, thanks to the good spatial resolution along the three axes 
\cite{billard.track}.  Nonetheless, electron/recoil discrimination remains a key and challenging issue. 
While a high rejection power is expected at high energy (above 20 keV), a dedicated data
analysis is needed at low energy. After identifying discriminant observables, a multivariate analysis, namely a Boosted Decision Tree, is 
proposed, enabling an efficient event tagging for Dark Matter search. We show that it allows us to optimize  rejection while keeping 
a rather high efficiency which is compulsory for rare event search.With respect to a  sequential analysis, the 
rejection is about $\sim 20$ times higher with a multivariate analysis, for the same Dark Matter exclusion limit. This event tagging strategy is of general interest for Dark Matter search and could be
adapted for various e/recoil discrimination strategies.


\section*{References}
 

\begin{thebibliography}{10}


\bibitem{spergel}D.~N.~Spergel, 
 Phys.\ Rev.\  D {\bf 37} (1988) 1353

 



\bibitem{albornoz}
D. Albornoz V\'asquez, G. B\'elanger, J. Billard and F. Mayet, Phys. Rev. D {\bf 85} (2012) 055023 


\bibitem{Alves:2012ay}
  D.~S.~M.~Alves, S.~E.~Hedri and J.~G.~Wacker,
  arXiv:1204.5487 [astro-ph.GA].

\bibitem{billard.disco}J.~Billard, F.~Mayet, J.~F.~Macias-Perez and D.~Santos, Phys. Lett. B {\bf 691} (2010) 156-162
\bibitem{billard.ident}J.~Billard, F.~Mayet and D.~Santos, Phys.\ Rev.\  D {\bf 83 } (2011)  075002




\bibitem{billard.profile}J.~Billard, F.~Mayet and D.~Santos,  Phys.\ Rev.\  D {\bf 85} (2012) 035006
\bibitem{billard.exclusion}
J.~Billard, F.~Mayet and D.~Santos,
 Phys.\ Rev.\  D {\bf 82} (2010) 055011 



\bibitem{Bozorgnia:2011vc}
  N.~Bozorgnia, G.~B.~Gelmini and P.~Gondolo,
  arXiv:1111.6361 [astro-ph.CO].

\bibitem{Chiang:2012ze}
  C.~-T.~Chiang, M.~Kamionkowski and G.~Z.~Krnjaic,
  arXiv:1202.1807 [astro-ph.CO].


\bibitem{copi1}C.~J.~Copi and L.~M.~Krauss,, Phys.\ Rev.\  D {\bf 63} (2001) 043507
\bibitem{copi2}C.~J.~Copi, L.~M.~Krauss, D.~Simmons-Duffin and S.~R.~Stroiney, Phys.\ Rev.\  D {\bf 75} (2007) 023514,   
\bibitem{copi3}C.~J.~Copi and L.~M.~Krauss,   Phys.\ Lett.\  B {\bf 461} (1999) 43

\bibitem{Creswick:2010dm}
  R.~J.~Creswick, S.~Nussinov and F.~T.~Avignone,
  Astropart.\ Phys.\  {\bf 35} (2011) 62


\bibitem{Kuhlen:2012fz}
  M.~Kuhlen, M.~Lisanti and D.~N.~Spergel,
  arXiv:1202.0007 [astro-ph.GA].

\bibitem{Lee:2012pf}
  S.~K.~Lee and A.~H.~G.~Peter,
  JCAP {\bf 1204} (2012) 029


  

\bibitem{Lisanti:2009vy}
  M.~Lisanti and J.~G.~Wacker,
  Phys.\ Rev.\ D {\bf 81} (2010) 096005


\bibitem{green1}A.~M.~Green and B.~Morgan, Phys.\ Rev.\ D {\bf 77}  (2008) 027303
\bibitem{green2}A.~M.~Green and B.~Morgan, Astroparticle Physics 27  (2007) 142
\bibitem{green.disco}A.~M.~Green and B.~Morgan, Phys.\ Rev.\  D {\bf 81} (2010) 061301

\bibitem{morgan1}B.~Morgan and A.~M.~Green, Phys.\ Rev.\ D {\bf 72}  (2005) 123501
\bibitem{morgan2}B.~Morgan, A.~M.~Green and N.~J.~C.~Spooner, Phys.\ Rev.\ D {\bf 71}  (2005) 103507


\bibitem{white}S.~Ahlen {\it et al.}, Int.\ J.\ Mod.\ Phys.\  A {\bf 25} (2010) 1



\bibitem{dmtpc}S.~Ahlen {\it et al.},
  Phys.\ Lett.\  {\bf B695 } (2011)  124-129
  
  
\bibitem{drift}
  E.~Daw {\it et al.}, Astropart.Phys. {\bf 35} (2012) 397-401 

\bibitem{d3}S.~E.~Vahsen {\it et al.},  EAS Publications Series {\bf 53} (2012) 43-50   

\bibitem{mimac}D.~Santos {\it et al.}, EAS Publications Series {\bf 53} (2012) 25-31

 
 
\bibitem{newage}K.~Miuchi {\it et al.}, Phys.\ Lett.\  B {\bf 686} (2010) 11


  
 
\bibitem{censier}B.~Censier, EAS Publications Series 53 (2012) 59-66 


\bibitem{cdms}Z.~Ahmed {\it et al.}, Phys.\ Rev.\ Lett.\  {\bf 106} (2011) 131302
\bibitem{edelweiss}E.~Armengaud {\it et al.}, Phys.\ Lett.\ B {\bf 702} (2011) 329
\bibitem{xenon1}E.~Aprile {\it et al.}, Astropart.\ Phys.\  {\bf 35} (2012) 573
\bibitem{xenon2}E.~Aprile {\it et al.}, Phys. Rev. D. {\bf 72} (2005) 072006
\bibitem{cresst}G.~Angloher  {\it et al.}, arXiv:1109.0702 [astro-ph.CO].

\bibitem{moulin.these} E.~Moulin,  PhD Thesis,  Universit\'e Joseph Fourier, Grenoble (France), Sept.
2005

 \bibitem{Santos:2005xj}D.~Santos, E.~Moulin, F.~Mayet and J.~Macias-Perez, J.\ Phys.\ Conf.\ Ser.\  {\bf 39} (2006) 154


 \bibitem{billard.track}J.~Billard, F.~Mayet and D.~Santos, JCAP {\bf 04} (2012) 006

 \bibitem{Giomataris1}Y.~Giomataris {\it et al.}, Nucl. Instrum. Meth. A  {\bf 560} (2006) 405   

 


\bibitem{Iguaz:2011yc}F.~J.~Iguaz {\it et al.}, JINST {\bf 6 } (2011)  P07002
  
   
  
\bibitem{Richer:2009pi}J.~P.~Richer {\it et al.}, Nucl.\ Instrum.\ Meth.\  A {\bf 620} (2010) 470
 
\bibitem{Bourrion1}O. Bourrion {\it et al.}, Nucl. Instrum. Meth. A {\bf 662} (2010) 207  
\bibitem{Bourrion2}O. Bourrion {\it et al.},  EAS Publications Series {\bf 53} (2012) 129-136
  
 
 
\bibitem{guillaudin}O. Guillaudin {\it et al.}, EAS Publications Series {\bf 53} (2012) 119-127
\bibitem{billard.these}J.~Billard, PhD Thesis,  Universit\'e Joseph Fourier, Grenoble (France), June 2012
\bibitem{magboltz}S. F.  Biagi,  Nucl. Instrum. and Meth. A {\bf 421} (1999) 234-240



\bibitem{g4}S. Agostinelli, Nucl.\ Instrum.\ Meth.\  A {\bf 506} (2003) 250-303
\bibitem{srim}J. F. Ziegler,  J. P. Biersack and U. Littmark U.,  Pergamon Press New York, (1985), www.srim.org.
\bibitem{miniboone}B.~P.~Roe {\it et al.}, Nucl. Instrum. Meth. A {\bf 543} (2005) 577-584
\bibitem{d0.top}V.~M.~Abazov {\it et al.}  Phys.\ Rev.\ D {\bf 78} (2008) 012005
\bibitem{Conrad:2006ip}J.~Conrad and F.~Tegenfeldt, JHEP {\bf 0607} (2006) 040
  
     
 
\bibitem{bdt}Y.~Coadou, EPJ Web of Conferences {\bf 4} (2010) 02003, Proceedings of SOS'08, IN2P3 School of Statistics, A.~Lucotte \& F.
Melot (Eds.)  


\bibitem{tmva} A. Hoecker {\it et al.}, PoS ACAT 040 (2007) 

 \bibitem{Lopez:2011yv}
 J.~P.~Lopez {\it et al.},
 arXiv:1109.3501 [astro-ph.IM].


\end{thebibliography}
\end{document}